\documentclass[sigconf]{acmart}

\usepackage[english]{babel}

\usepackage[noend]{algpseudocode}
\usepackage{dsfont}
\usepackage{bbm}
\usepackage[subtle]{savetrees}
\usepackage[small,compact]{titlesec}
\usepackage{filecontents}
\usepackage{xcolor}
\usepackage{xspace}
\usepackage{mathtools}
\usepackage{amsmath}
\usepackage{amsthm}
\usepackage{subcaption}
\usepackage{makecell}
\usepackage{wrapfig}
\usepackage{xurl}
\usepackage{booktabs}
\usepackage{natbib} 
\usepackage{enumitem}
\usepackage{graphics}
\usepackage{graphicx}
\usepackage{grffile}
\usepackage{siunitx}
\usepackage{tikz}
\usepackage{xcolor}
\usepackage{tabulary}
\usepackage[font={small}]{caption}

\renewcommand\footnotetextcopyrightpermission[1]{} 
\setcopyright{none}

\acmDOI{}

\acmISBN{}

\acmPrice{}

\begin{document}
\title{\TheSystem: Efficient and Responsive Rate Control \\ for Real-Time Video on Variable Networks}
\renewcommand{\shorttitle}{Efficient and Responsive Rate Control for Real-Time Video}



\author{
\quad \\
\textnormal{Pantea Karimi
\hspace{0.3em} Sadjad Fouladi~\raisebox{0.3ex}{\small$\boxplus$}
\hspace{0.3em} Vibhaalakshmi Sivaraman 
\hspace{0.3em} Mohammad Alizadeh} \\
\quad \\
\textit{Massachusetts Institute of Technology}, \raisebox{0.3ex}{\small$\boxplus$}~\textit{Microsoft Research}
\vspace{3ex}
\quad \\
}


\renewcommand{\shortauthors}{Karimi et al.}
\newcommand*\circled[1]{\tikz[baseline=(char.base)]{
            \node[shape=circle,line width=0.75pt, draw=black,inner sep=1pt] (char) {\textcolor{black}{\small\textsf{#1}}};}}
\newcommand*\sfboxed[1]{\tikz[baseline=(char.base)]{
            \node[shape=rectangle,line width=0.75pt, draw=black,inner sep=2pt, rounded corners=2pt] (char) {\textcolor{black}{\small\textsf{#1}}};}}

\newcommand{\eg}{{\em e.g., }}
\newcommand{\etal}{{\em et al.}}
\newcommand{\etc}{{\em etc.}}
\newcommand{\vs}{{\em vs. }}
\newcommand{\ie}{{\em i.e., }}
\newcommand{\cwnd}{{\em cwnd}\xspace}
\newcommand{\fr}{frame rate\xspace}
\newcommand{\fqd}{frame queueing delay\xspace}
\newcommand{\fps}{FPS\xspace}
\newcommand{\pst}{pacer service time\xspace}
\newcommand{\fst}{frame service time\xspace}
\newcommand{\fsts}{frame service times\xspace}
\newcommand{\inflight}{{\em inflight }}
\newcommand{\Sec}[1]{\S\ref{#1}}
\newcommand{\SubSec}[1]{\S\ref{#1}}
\newcommand{\App}[1]{App.~\ref{app:#1}}
\newcommand{\Alg}[1]{Algorithm~\ref{alg:#1}}
\newcommand{\TheSystem}{Vidaptive\xspace} 
\newcommand{\vce}{video conference\xspace}
\newcommand{\vces}{video conferences\xspace}
\newcommand{\vcing}{video conferencing\xspace}
\newcommand{\Vcing}{Video conferencing\xspace}
\newcommand{\bpp}{bits-per-pixel\xspace}
\newcommand{\perframe}{per-frame\xspace}
\newcommand{\Fig}[1]{Fig.~\ref{fig:#1}}
\newcommand{\NewPara}[1]{\noindent{\bf #1}}
\newcommand{\NewParaI}[1]{\noindent{\emph{#1}}}
\newcommand{\eat}[1]{}
\newcommand{\Tab}[1]{Tab.~\ref{tab:#1}\xspace}
\newcommand{\Eqn}[1]{Eq.~\ref{eqn:#1}\xspace}
\newcommand{\Thm}[1]{Theorem.~\ref{thm:#1}}
\newcommand{\Cor}[1]{Imp.~\ref{cor:#1}}
\newcommand{\cc}{congestion control\xspace}
\newcommand{\rocc}{RoCC\xspace}
\newcommand{\cca}{CCA\xspace}
\newcommand{\ccas}{CCAs\xspace}
\newcommand{\CC}{Congestion Control\xspace}
\newcommand{\alg}{algorithm\xspace}
\newcommand{\ack}{acknowledgment\xspace}
\newcommand{\ccrate}{{\tt CC-Rate}\xspace}
\newcommand{\CCA}{Congestion Control Algorithm\xspace}
\newcommand{\VCCA}{Video Congestion Control Algorithm\xspace}
\newcommand{\vcca}{video congestion control algorithm\xspace}

\begin{abstract}
        \noindent Real-time video streaming relies on rate control mechanisms to adapt video bitrate to network capacity while maintaining high utilization and low delay. However, the current video rate controllers, such as Google Congestion Control (GCC), are very slow to respond to network changes, leading to link under-utilization and latency spikes. While recent delay-based congestion control algorithms promise high efficiency and rapid adaptation to variable conditions, low-latency video applications have been unable to adopt these schemes due to the intertwined relationship between video encoders and rate control in current systems. 

This paper introduces \TheSystem, a new rate control mechanism designed for low-latency video applications. \TheSystem decouples packet transmission decisions from encoder output, injecting ``dummy'' padding traffic as needed to treat video streams akin to backlogged flows controlled by a delay-based congestion controller. \TheSystem then adapts the target bitrate of the encoder based on delay measurements to align the video bitrate with the congestion controller's sending rate. Our evaluations atop Google's implementation of WebRTC show that, across a set of cellular traces, \TheSystem achieves $\sim$1.5$\times$ higher video bitrate and \SI{1.4}{dB} higher SSIM, \SI{1.3}{dB} higher PSNR, and 40\% higher VMAF, and it reduces $95^\text{th}$-percentile frame latency by \SI{2.2}{s} with a slight \SI{17}{ms} increase in median frame latency.

\vspace{-5 pt}
\end{abstract}

\maketitle

\section{Introduction}
\label{sec:intro}

Real-time video streaming has become an integral part of modern communication systems, enabling various applications from video conferencing to cloud gaming, live video, and teleoperation. A critical component of these systems is the rate control mechanism, which adapts the video bitrate to the available network capacity. State-of-the-art rate controllers, however, such as Google Congestion Control\footnote{GCC is the default rate controller in Google's implementation of WebRTC, which is deployed across billions of devices~\cite{Bubley2015WebRTC}. Since the publication of the GCC paper, its algorithm has evolved. We use the most recent version in our research.} (GCC)~\cite{gcc}, have significant shortcomings. Specifically, GCC is slow to adapt to changes in network conditions, leading to both link under-utilization and latency spikes. 

In recent years, several congestion control algorithms (CCA) have been proposed that achieve high utilization, low delay, and fast convergence~\cite{arun2018copa, cardwell2016bbr, sprout, verus}. These algorithms are highly responsive to network variations, adapting within a few round-trip times (RTTs). In contrast, video rate controllers like GCC lag considerably. They can take an order of magnitude longer than state-of-the-art CCAs to increase the video bitrate when extra bandwidth becomes available, while also reacting slowly to sudden bandwidth drops. The slow reaction time significantly degrades performance in variable networks such as wireless links~\cite{sprout}. In our experiments using cellular network traces, GCC under-utilizes the network by 3$\times$ compared to Copa~\cite{arun2018copa} while incurring 7$\times$ higher tail (P95) packet latency on average. 

This sluggishness is not merely a limitation of the design of today's video rate controllers but a symptom of a broader issue: the inherent coupling between the video encoder and rate control~\cite{salsify}. Current systems rely on {\em encoder-driven} rate control mechanisms that adapt the video bitrate by controlling the encoder's target bitrate. Each video frame is sent almost immediately and in full when it arrives, so the instantaneous rate ``on the wire'' is dictated by the size of the frames produced by the encoder. However, standard video encoders are not designed to adjust to rapid changes in network conditions and generally take several frames to adapt the frame size to a new target bitrate~\cite{salsify}. Moreover, the frame sizes are variable and only meet the target bitrate on average in a best-effort manner. To maintain low latency despite the uncertainties of the encoder, algorithms like GCC set the target bitrate conservatively and increase it slowly. Nevertheless, during times of significant congestion (e.g., due to link outages), the encoder cannot immediately adapt, and GCC experiences significant latency spikes.

Recently, Salsify~\cite{salsify} introduced a responsive architecture for real-time video systems by proposing a design centered around a ``functional'' video codec. This design allows the encoder to adapt more quickly to rate control decisions but requires a new interface to the video codec.
Due to the ubiquity of codecs embedded in a wide array of devices, from smartphones to tablets and personal computers to smart TVs, implementing such widespread changes to video codecs presents a substantial challenge, requiring not only software updates but potentially also hardware revisions~\cite{6737724, bitmovin2023, NVIDIAVideoCodecSDK}. Moreover, our experiments show Salsify can still benefit from more efficient rate control, as its default rate control mechanism leads to increased tail latency in highly variable links.

We present \TheSystem, a new rate control mechanism for low-latency video applications that significantly improves efficiency and responsiveness to network variability without needing encoder modifications. \TheSystem's design is based on two key concepts.

The first is to decouple instantaneous packet transmission decisions from the encoder's output. Specifically, \TheSystem treats video streams as if they were backlogged flows for the purpose of rate control, allowing it to use any delay-based CCA to control the sending rate (we primarily use Copa~\cite{arun2018copa} in our implementation). If the encoder produces more packets than the CCA is willing to send, it buffers them at the sender and paces them out at the CCA's chosen rate. On the other hand, \TheSystem sends ``dummy packets'' to fill the gap if the encoder does not produce enough packets to sustain the CCA's rate. This approach ensures that, on the wire, \TheSystem behaves identically to its adopted CCA running with a backlogged flow. Since the CCA utilizes the link at or near capacity at all times, its feedback loop can operate without disruption and track the available bandwidth quickly.

\TheSystem's second concept is a set of techniques to match the video bitrate to the CCA's sending rate to the extent possible while controlling frame latency. To keep the frame latency within acceptable bounds, \TheSystem skips a frame if the delay through the sender's queue exceeds a threshold (e.g., when the encoder overshoots the CCA's rate). This momentarily reduces the frame rate to handle sudden latency spikes. Further, \TheSystem uses an online algorithm to determine the encoder's target bitrate based on the CCA's sending rate and recent frame transmission times. The algorithm continually measures recent frame transmission times out of the sender's queue, and calculates a {\em headroom} between the encoder's target bitrate and the CCA's rate that is needed to achieve a desired frame delay. By considering the actual distribution of observed frame delays, \TheSystem automatically adapts the headroom to the variability in the encoder's output and the network rate. It provides simple knobs to navigate the tradeoff between video bitrate and frame latency, enabling different operating points depending on the application. For example, in interactive applications like video conferencing, \TheSystem can be tuned for ultra-low latency while a live streaming application can improve video quality for a modest increase in latency.

We implement \TheSystem atop Google's WebRTC (version M108~\cite{webrtc_git_commit}) and test it using fifteen cellular network traces with significant variability and twenty 1080p YouTube videos from diverse genres on emulated network links~\cite{netravali2015mahimahi}. Compared to GCC, our key findings are across all videos and all traces:
\begin{enumerate}[noitemsep,topsep=0pt,parsep=0pt,partopsep=0pt]
    \item \TheSystem improves link utilization by $\sim$3$\times$ and video bitrate by $\sim$1.5$\times$ on average.
    \item \TheSystem achieves \SI{1.4}{dB} improvement in Structural Similarity Index Measure (SSIM), 40\% increase in Video Multimethod Assessment Fusion (VMAF), and \SI{1.3}{dB} in Peak-Signal-to-Noise-Ratio (PSNR) on average. 
    \item  \TheSystem reduces $95^\text{th}$ percentile frame latency by \SI{2246}{ms}. \TheSystem increases median latency by a slight \SI{17}{ms}.
    \item \TheSystem reduces frame rate by $\sim$ 10\% on average.
\end{enumerate}
We also ran experiments on several real-world calibrated Pantheon traces and observed similar results~\cite{pantheon}.

Compared to Salsify, \TheSystem achieves \SI{6.3}{s} better P95 latency and \SI{52}{ms} better P50 latency across all the videos and traces. Salsify's default rate control mechanism transmits at a higher rate than \TheSystem, resulting in better video quality (e.g., 35\% higher VMAF score), but it's unable to control the latency.  

\vspace{-5 pt}
\paragraph{Ethics Statement.} This paper does not raise any ethical concerns or issues.

\section{Motivation and Key Ideas}
\label{sec:motivation}

\begin{figure*}[t]
\centering
    \begin{subfigure}{\linewidth}
        \centering
        \includegraphics[width=\linewidth]{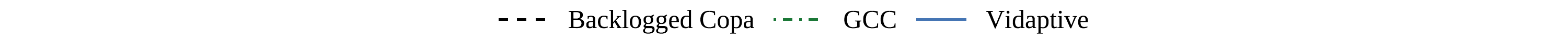}
        \vspace{-10pt}
        \label{fig:motivation_legend}
    \end{subfigure}
    \begin{subfigure}{0.3\linewidth}
        \centering
        \includegraphics[width=\linewidth]{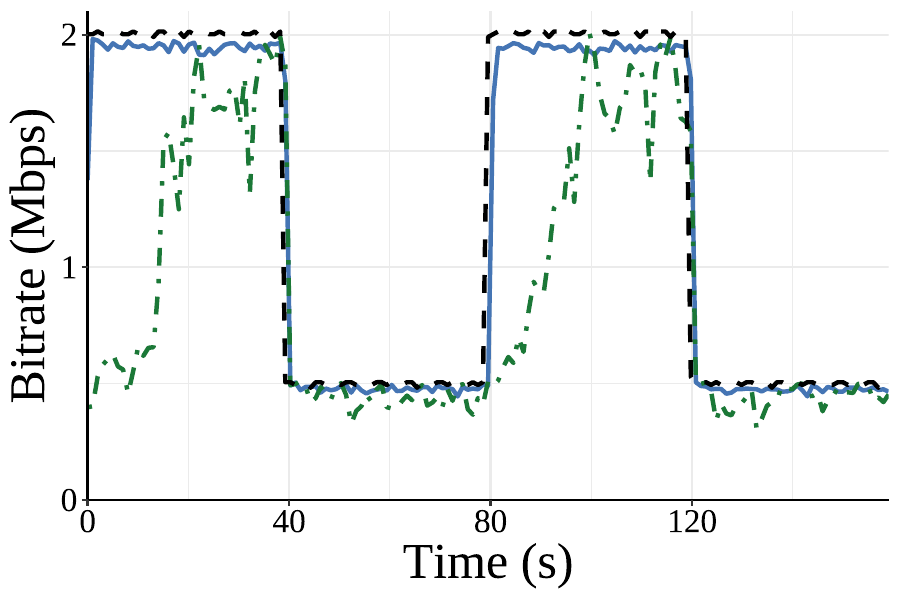}
        \caption{\small Link Utilization}
        \label{fig:motivation_bitrate}
    \end{subfigure}
    \begin{subfigure}{0.3\linewidth}
        \centering
        \includegraphics[width=\linewidth]{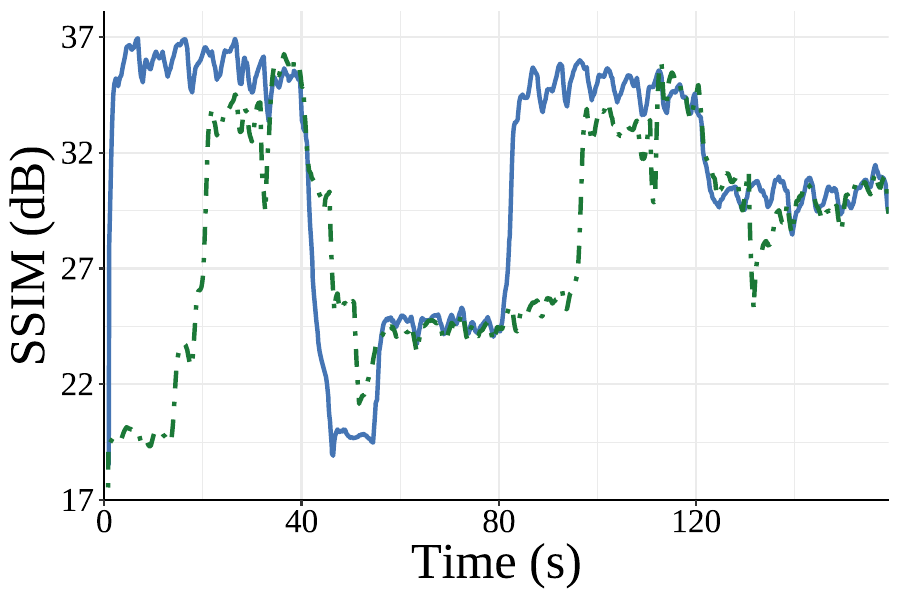}
        \caption{\small Frame Quality}
        \label{fig:motivation_ssim}
    \end{subfigure}
    \begin{subfigure}{0.3\linewidth}
        \centering
        \includegraphics[width=\linewidth]{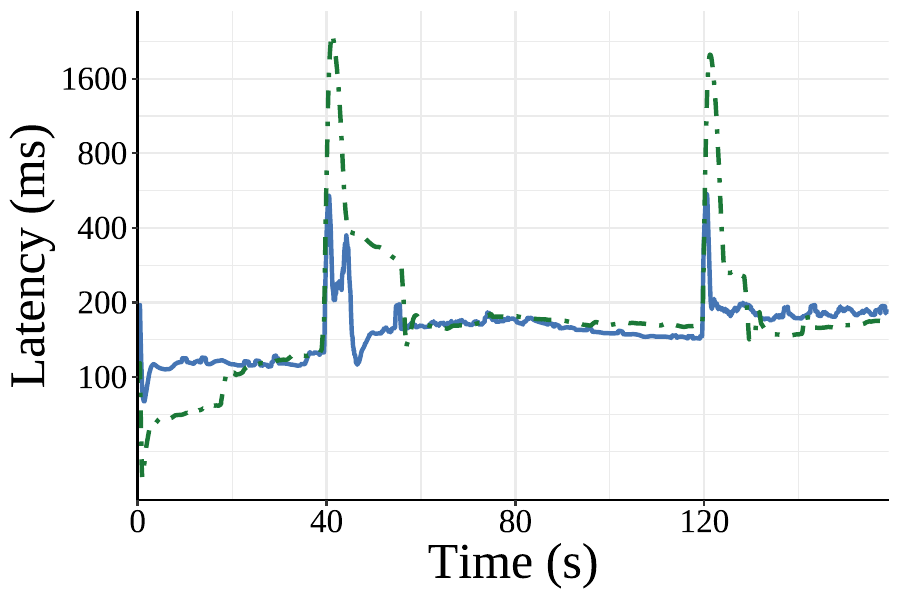}
        \vspace{-5 pt}
        \caption{\small Frame Latency}
        \label{fig:motivation_latency}
    \end{subfigure}
    \caption{\small Utilization, frame quality, and latencies of Copa on a backlogged flow, GCC on a video flow, and Copa + \TheSystem on a video flow. GCC is very slow to match the available capacity and under-utilizes the link in the steady state. 
    Copa + \TheSystem responds much faster to link variations and is similar to Copa's performance on a backlogged flow. }
\vspace{-10pt}
\label{fig:motivation}
\end{figure*}

\subsection{The Problem}
\label{sec:motivation_problem}
\NewPara{Status Quo for Video Rate Control.}
To understand how rate control for real-time video works today, we run Google Congestion Control (GCC)~\cite{gcc}, the rate control mechanism inside WebRTC, on an emulated link that alternates between \SI{2}{Mbps} and \SI{500}{Kbps} every 40 seconds. We use a recent version of WebRTC (M108~\cite{webrtc_git_commit}) in all our experiments. The minimum network round-trip time (RTT) is \SI{50}{ms}, and the buffer size at the bottleneck is large enough that there are no packet drops.
In \Fig{motivation_bitrate}, GCC is sluggish to increase its rate when the stream begins and when the link capacity rises back to \SI{2}{Mbps} at $t= 80$s. Specifically, GCC takes \emph{18 seconds} to go from \SI{500}{Kbps} to \SI{2}{Mbps}, resulting in lower visual quality during that time (\Fig{motivation_ssim}). GCC is also slow to react to capacity drops: in \Fig{motivation_latency}, when the link rate drops to \SI{500}{Kbps} at $t=40$s, GCC's frame latency spikes to over a second and settles only after 12 \emph{seconds}.

Contrast this behavior with traditional congestion control  algorithms~\cite{arun2018copa,rocc,vegas,cardwell2016bbr} operating on \emph{backlogged} flows: they respond to such network events much faster, typically over few RTTs.
For instance, the ``Backlogged Copa'' lines in \Fig{motivation_bitrate} and \Fig{motivation_latency} show that Copa~\cite{arun2018copa},
running on a backlogged flow on the same time-varying link, is much more responsive to the network conditions.
This wide disparity between GCC and Copa raises the question: {\em Why does state-of-the-art video rate control lag so far behind state-of-the-art congestion control?}

\begin{figure}
\centering
    \includegraphics[width=0.8\linewidth]{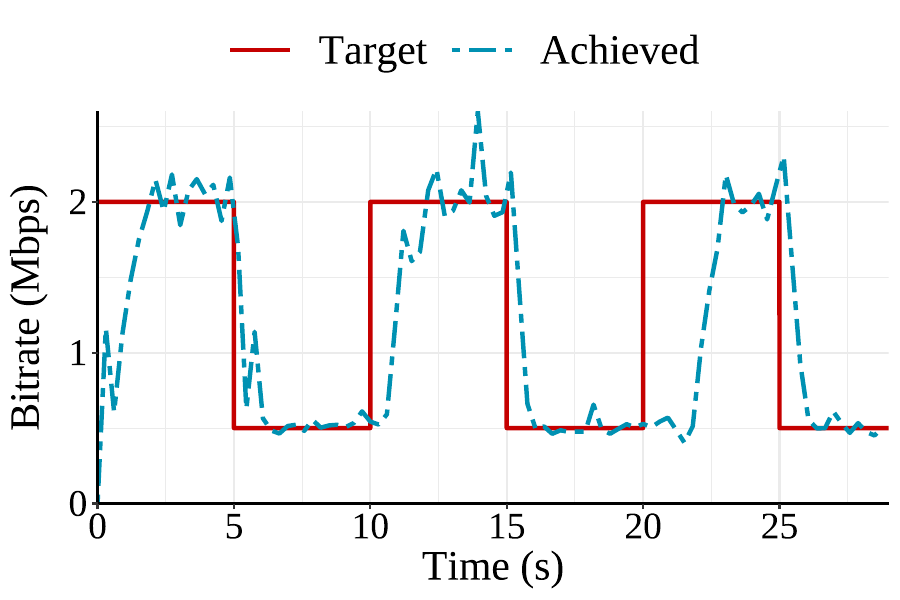}
    \vspace{-9 pt}
    \caption{\small Video encoder's response to a time-varying ``Target'' input bitrate. ``Achieved'' reflects the encoder's output rate. The encoder is slow to increase its output rate and exhibits lots of variation around the average output rate in the steady state.
    }
\vspace{-20pt}
\label{fig:encoder reaction time}
\end{figure}

\NewPara{Encoder-driven Rate Control Loop.}
In video rate controllers like GCC, the {\em instantaneous rate} at which data is transmitted on the wire is dictated by the size of the video frames produced by the encoder. GCC seeks to control the video bitrate by adapting the encoder's target rate, but the encoded frame sizes can be highly variable. The encoder achieves its target bitrate only on average---usually throughout several frames~\cite{salsify}. Moreover, the encoder's bitrate cannot immediately adapt to changes in the target bitrate. We illustrate this behavior in \Fig{encoder reaction time} where we supply the VP8 encoder with a target bitrate that switches between \SI{2}{Mbps} and \SI{500}{Kbps} every 5s and observe its achieved bitrate. Every time the bitrate goes up from \SI{500}{Kbps} to \SI{2}{Mbps}, the encoder takes nearly 2 seconds to catch up.
On the way down from \SI{2}{Mbps} to \SI{500}{Kbps}, it takes about a second to lower the bitrate. 
The reason for this lag is that the size of an encoded frame depends on several factors, including quantization parameters, the encoder's internal state, and the motion in the video, and it is only known accurately after encoding. The encoder tries to rectify its over- and under-shootings by adjusting the quality of subsequent frames. However, even after the encoder matches the target bitrate, it exhibits considerable variance around the average on a per-frame basis.

Since the encoder cannot match the target bitrate on per-frame timescales, GCC cannot immediately reduce the bitrate if the capacity suddenly drops. Instead, GCC has to be conservative and leave abundant bitrate {\em headroom} at all times. Even in the steady state, GCC limits utilization not to exceed 85\%, and it experiences occasional latency spikes (\Fig{motivation_latency}). Headroom reduces the risk of a latency spike during fluctuations but makes it difficult to track the available bandwidth accurately. For example, GCC uses delay gradients to estimate the link rate but if the link is significantly under-utilized, there is almost no queuing and delay gradients do not provide a signal. Thus GCC must slowly ramp up its target bitrate rate to discover capacity. Recall that when GCC increases its target bitrate, the encoder takes a few seconds increase the video bitrate. GCC must wait for the video bitrate to increase before it knows if it is safe to increase the target bitrate further, so the entire process takes 15--20 seconds to converge (\Fig{motivation_bitrate}).

\NewPara{Basic probing mechanisms are not enough.} Video systems like WebRTC support basic mechanisms~\cite{webrtc-probing} that send padding to probe for extra bandwidth. For example, GCC fires a periodic timer (e.g., every \SI{5}{s}) and sends some extra padding traffic to check for bandwidth. This coarse probing cannot provide precise rate control, which requires RTT timescale feedback. \Fig{probing} in Appendix~\ref{sec:probing_appendix} shows that GCC has a similar behavior on the above alternating link with and without probing.

\subsection{Our Solution}
\label{sec:motivation_our_solution}
\NewPara{Decoupling the Encoder from the Rate on the Wire.}
As illustrated in \Fig{motivation_bitrate}, a backlogged flow using a state-of-the-art \cca, like Copa, can adapt to time-varying network capacity on RTT timescales while also controlling network queueing delay. A key reason is that such \ccas have fine-grained control over when to send each packet, e.g., driven via the ``ACK clock''~\cite{jacobson1988congestion}. \TheSystem makes video streams appear like a backlogged flow to the congestion controller. This allows \TheSystem to leverage existing \ccas optimized for high throughput, low delay, and fast convergence. In \Fig{motivation}, \TheSystem using Copa for congestion control achieves nearly identical throughput and network delay as a backlogged Copa flow. \TheSystem quickly increases its bitrate when bandwidth increases, leading to higher image quality than GCC following each such event (\Fig{motivation_ssim}) while having lower latency spikes than GCC (\Fig{motivation_latency}).

\TheSystem sends packets on the wire as dictated by the congestion controller. Specifically, when the encoder overshoots the available capacity, \TheSystem queues excess video packets in a buffer and only sends them out when congestion control allows (e.g., according to the congestion window and in-flight packets for window-based \ccas). Conversely, when the encoder undershoots the available capacity, \TheSystem sends ``dummy packets'' to match the rate requested by the congestion control by padding the encoder output with additional traffic.\footnote{This dummy traffic could also be repurposed for helpful information such as forward error correction (FEC) packets~\cite{rudow2023tambur, nagy2014congestion, johanson2003adaptive} or keyframes for faster recovery from loss. We leave such enhancements to future work and focus solely on the impact of dummy traffic on video congestion control.}
This allows the \cca to operate without disruption (as in a backlogged flow) despite the encoder's varying frame sizes. Our evaluations show that, unlike the common belief that sending dummy traffic ``seldom works for real-time video''~\cite{onrl}, carefully using dummy traffic within our design improves the video bitrate while having minimal effect on the latency.

By decoupling the congestion controller's decisions from the encoder output, \TheSystem can accurately track time-varying bottleneck rates. However, it is still important to match the actual video bitrate produced by the encoder to the congestion controller's sending rate. In particular, although buffering packets and sending dummy traffic can handle brief variations in the encoder output bitrate, the quality of experience will suffer if the encoder's output is persistently higher or lower than the CCA's rate. In the former case, end-to-end frame latency would grow uncontrollably, and in the latter scenario, the opportunity to have a better quality is missed. \TheSystem includes mechanisms that control the encoder's target bitrate to match the video bitrate to the CCA's sending to the extent possible while controlling frame latency, as discussed in \S\ref{sec:design}.

\section{\TheSystem Design}
\label{sec:design}
\subsection{Overview}

\begin{figure*}
\centering

\begin{minipage}[t]{0.64\textwidth}
\vspace{0pt}
    \includegraphics[width=\linewidth]{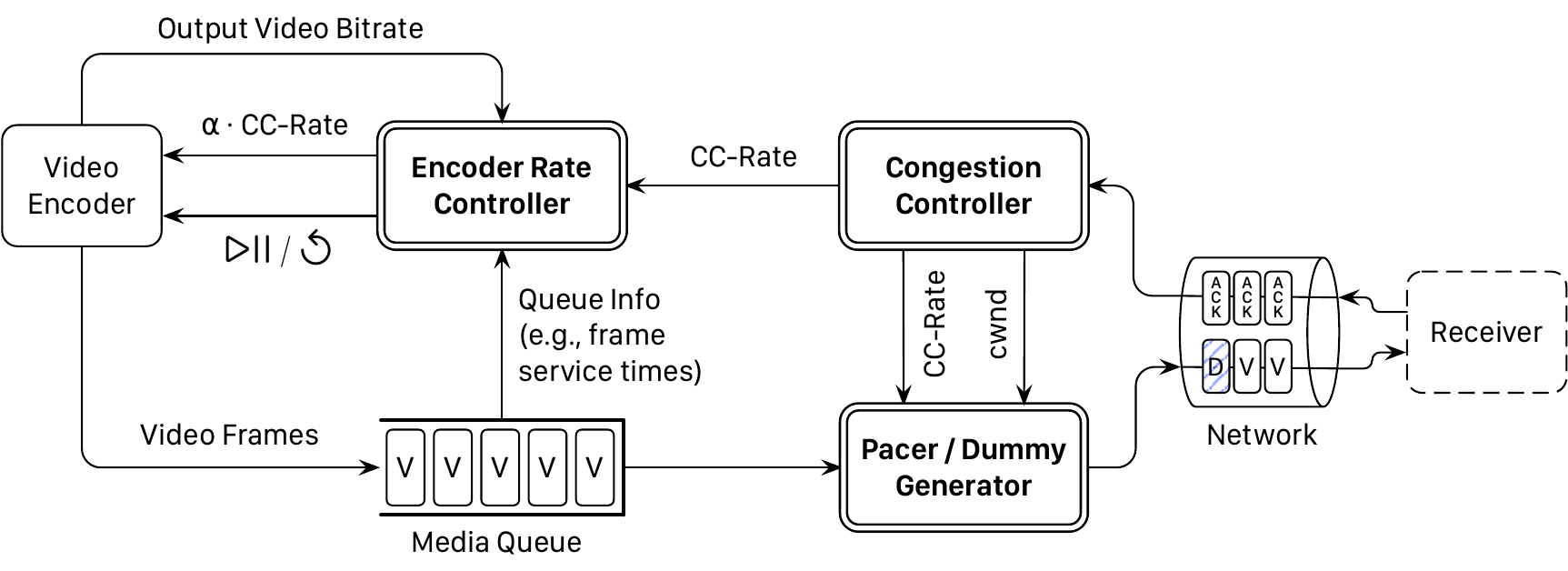}
\end{minipage}\hfill
\begin{minipage}[t]{0.33\textwidth}
\setlength{\abovecaptionskip}{0pt}
\vspace{0pt}
\caption{\small \TheSystem Design. \TheSystem uses a window-based Congestion Controller, Pacer, and a new Dummy Generator to decouple the rate at which traffic is sent on the wire from the encoder. The Encoder Rate Controller monitors frame delays to trigger latency safeguards and picks a new target bitrate based on the discrepancy between the \ccrate and the video encoder's current bitrate.}\label{fig:design}
\end{minipage}
\end{figure*}

\Fig{design} shows \TheSystem's overall design. The video encoder encodes frames and sends them to an application-level media queue before sending the packets into the network. At the transport layer, we add a window-based ``Congestion Controller,'' a ``Pacer,'' and a ``Dummy Generator'' to decouple the rate at which traffic is sent on the wire from what is produced by the encoder, as described in \Sec{sec:design transport}. We introduce an ``Encoder Rate Controller'' that monitors the queueing delay frames are experiencing to trigger the latency safeguards described in \Sec{sec:design latency}. The Encoder Rate Controller also controls the encoder's target bitrate based on the discrepancy between the CCA's sending rate, the video encoder's current bitrate, and the observed frame transmission times (\Sec{sec:design tradeoff}). 

\subsection{Transport Layer}
\label{sec:design transport}
\NewPara{Congestion Controller.} \TheSystem uses a delay-controlling \emph{window-based} congestion control algorithm to 
determine when to send packets. The congestion window (\cwnd) limits the maximum number of in-flight bytes between the sender and the receiver. 
The sender keeps track of the number of bytes in flight and only sends out new packets if the amount in-flight is less than \cwnd. \TheSystem can use any CCA that controls delay effectively. Our focus is not on designing a new CCA; fortunately, several excellent algorithms already exist. We use two recently proposed algorithms, Copa~\cite{arun2018copa} and RoCC~\cite{rocc} in our implementation. Most of our results use Copa, which keeps a tight check on the network delay, reacts to changes quickly, and maintains high utilization. \TheSystem's sending behavior is identical to its underlying CCA (\S\ref{sec:motivation_our_solution}), so it inherits these properties as well as the flow-level fairness characteristics of the CCA.

\NewPara{Pacer.} Beside enforcing the \cwnd constraint, the sender also paces the packet transmissions at the CCA's sending rate (\ccrate), computed as the \cwnd divided by smoothed RTT (a weighted moving average of RTT samples). Since the encoder's output bitrate may instantaneously overshoot the available capacity (\Sec{sec:motivation}), the Pacer is responsible for avoiding a sudden burst of packets. 

\NewPara{Dummy Packets.} Recall that our goal is to emulate the behavior of a backlogged flow to enable the \cca to track changes in available bandwidth quickly and accurately (\S\ref{sec:motivation_our_solution}). To this end, we generate ``dummy packets'' if the \cca is ready to send a packet, but no video packets are available in the Pacer. Dummy packets have strictly lower priority than video packets; they are only sent if the Pacer queue is empty and congestion control detects free bandwidth.  To limit spurious dummy packets, we do not send them if the time to the next expected frame is lower than a threshold, e,g, 25\% of the inter-frame time (\SI{8.25}{ms} for \SI{30}{\fps}) in our implementation. The benefit of this mechanism is that
if the network is soon to deteriorate, the dummy packets sent a few milliseconds before a frame will cause a higher queuing delay in the network. On the other hand, if the network rate increases, not sending packets for a few milliseconds will not noticeably slow down the congestion controller's reaction time. 
Lastly, we stop sending dummy packets if the video has reached its maximum achievable bitrate (\SI{12}{Mbps} in our experiments).\footnote{
Our analysis of the 1080p videos within our dataset showed that the peak bitrate typically reached was around \SI{12}{Mbps}, so for this collection of videos, \SI{12}{Mbps} represents the upper limit of bitrate. \label{why_12mbps}} Since the video bitrate cannot increase further, sending extra traffic to discover more available bandwidth would not be useful.

\subsection{Safeguarding against Latency Spikes}
\label{sec:design latency}

\NewPara{Encoder Pause.} \TheSystem monitors the time packets spend in the pacer queue before they are sent out. If the time spent by the oldest packet exceeds a certain pause threshold ($\tau$), we \emph{pause} encoding of frames,
and only resume encoding only once the pacer queue has drained. 
We set $\tau$ = \SI{33}{ms} by default in our implementation for a camera rate of \SI{30}{\fps}, thereby pausing encoding if packets from the previous frame are yet to be sent out. The intuition here is that there is no point in encoding a frame that would have to sit in the Pacer queue, waiting for a previous frame to finish transmission. Instead, we always encode and transmit fresh frames when they have a high chance of reaching the receiver with acceptable latency. Note that a high delay through the Pacer queue reflects congestion at the bottleneck link. If we ignore {\tt CC-Rate} and transmit the packets stuck in the Pacer queue (as currently implemented in WebRTC), they would still have to wait at the bottleneck link, and it would take longer to drain these packets when the link bandwidth improves.

\subsection{Trading off Latency and Quality}
\label{sec:design tradeoff}
Different real-time video applications exhibit varying tolerance levels to frame latency. For instance, cloud gaming demands exceptionally low latency, prioritizing rapid response. Conversely, live-streaming applications can afford higher latency without significantly impacting the user experience. To address these differing needs, \TheSystem offers a mechanism to trade off the frame latency and bitrate (and, consequently, the frame quality). It's essential to consider the frame transmission time to maintain a low frame latency. Smaller frames, resulting from a lower bitrate, can be transmitted more quickly. However, this often comes at the expense of video quality. 

We formalize the tradeoff between frame latency and quality as a control problem that picks a target bitrate for the encoder based on how much we prioritize achieving a lower latency. Specifically, \TheSystem picks $\alpha$, the fraction of the \ccrate to supply as the target bitrate to the encoder. When the latency is high, \TheSystem chooses a smaller $\alpha$ to create smaller frames so they can be transmitted through the network faster. When the latency is low, \TheSystem chooses a higher $\alpha$ to obtain higher quality frames.

\NewPara{Preliminaries.}
To estimate the frame transmission time, we measure the \fst, which is the time between when the first packet
of a frame arrives at the front of the pacer until its last packet is sent out
on the network. The \fst is a function of the \ccrate and frame sizes. As a result, it is impacted by fluctuations in both the encoder's output and in the \ccrate. These fluctuations are out of our control and can be viewed as exogenous ``noise'' impacting \fsts. However, we can influence the {\em expected} frame sizes by controlling the encoder's {\em target bitrate}. The crux of our method is to pick the target bitrate in a way that controls the distribution of \fsts. 
\TheSystem measures the \fst for each frame in the Pacer.

Assume there are $N$ frames over a time interval $T$ that experience service times denoted by $d_i$ for $i \in \{1, 2, ..., N\}$,
and they each have been encoded by a target bitrate of $tr_i$.
The distribution of $d_i$ gives us important information about the latency of frames. If the camera's \fr is $f_{max}$ (typically \SI{30}{\fps}), the frame interval is $\Delta = \frac{1}{f_{max}}$ (typically \SI{33}{ms}). For maximum efficiency, the $d_i$ values should be close to, but less than, $\Delta$, such that the last packet of a frame is transmitted just as the next frame is encoded. For example, if the $80^{th}$ percentile of the $d_i$ distribution is larger than $\Delta$, 20\% of the frames take more than $\Delta$ to serve. These frames increase the delay of subsequent frames and may trigger our latency safeguard (\S\ref{sec:design latency}), causing frames to be skipped.  If $d_i$ values have a long tail, video frames will experience high tail latency.

\NewPara{Choosing a Target Bitrate.} Assume that a target bitrate  $\alpha \cdot \text{\ccrate}$ is given to the encoder where $0 < \alpha < 1$.
A larger $\alpha$ increases frame size and its $d_i$. Since $d_i$ depends on $\alpha$, we denote it as $d_i(\alpha)$. 
Our goal is to find $\alpha$ such that:
\begin{align}
\label{eqn:optimization}
\text{Percentile}_{\lambda}(\{d_i(\alpha)\}_{i=1}^{N}) = P
 \\ \nonumber
\textrm{s.t.} \quad 0 < \alpha < 1
\end{align}
\noindent where $\lambda \in (0, 1)$ is the specific percentile targeted for latency control. Most real-time applications would use a high $\lambda$ (0.9, for instance) to control the tail latency. For instance, when $\lambda \sim 1$, the focus shifts to maximum latency control. Meanwhile, $P$ captures the preference balance between latency and quality. When $P$ is picked closer to $\Delta$, this algorithm drives the \fsts closer to $\Delta$, increasing the video bitrate to near capacity. This increases video quality at the cost of potentially higher latency. When $P$ is picked smaller than $\Delta$, for example, $\frac{\Delta}{2}$, the target bitrate values are chosen conservatively, resulting in lower latency and potentially lower quality.

    \NewPara{Finding $\alpha$.} \label{optimization} To choose $\alpha$, one would ideally want to solve the above equation over {\em future} frames. However, it is hard to model $d_i(\alpha)$ for future frames since these can depend on future video content (e.g., the extent of motion) and how \ccrate changes in the future. Instead, we choose $\alpha$ based on the observed service times of {\em recent frames}. Assume we have \fsts $d_i$ for $i \in \{1, 2, .., N\}$ over the last $T$ second interval. We ask what value of $\alpha$ would have achieved Eq.~\eqref{eqn:optimization} for these frames? 

    Let $tr_i$ be the target bitrate used to encode frame $i$. Had frame $i$ been encoded by $\alpha \cdot \text{\ccrate}$ instead, the \fst would have been $\Tilde{d_i}(\alpha) = d_i \frac{\alpha \cdot \text{\ccrate}}{tr_i}$. We refer to $\Tilde{d_i}(\alpha)$ as the {\em counterfactual} service time for frame $i$.   
    This estimate assumes that frame size is proportional to the target bitrate (and hence proportional to $\alpha$) and that changing the target bitrate would not have changed \ccrate.\footnote{This holds because \TheSystem decouples \ccrate from the encoder.} Using these counterfactual estimates, we can now find $\alpha$ in Eq.~\eqref{eqn:optimization}. 

    \begin{align}
    \label{eqn:optimization_solution}
     \alpha = \min\Big(\frac{P}{\text{Percentile}_{\lambda}\big( \{d_i \frac{\text{\ccrate}}{tr_i}\}_{i=1}^{N} \big)}, 1\Big)
     \\ \nonumber
    \end{align}

\begin{figure}[t]
 \centering
 \begin{subfigure}[b]{0.325\linewidth}
    \vspace{-5 pt}
     \centering
     \includegraphics[width=\linewidth]{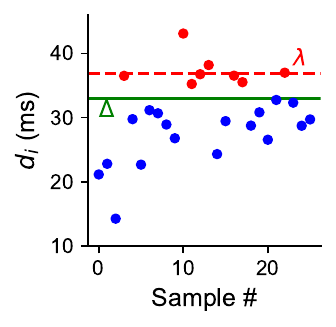}
     \caption{\centering  \small Service time samples}
     \label{fig:optimization_solution_before}
 \end{subfigure}
 \hfill
 \begin{subfigure}[b]{0.325\linewidth}
     \centering
     \includegraphics[width=\linewidth]{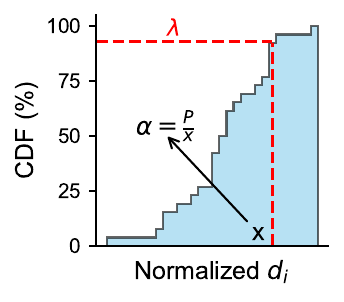}
     \caption{\centering \small 
      Solving the equation for $\alpha$
     }
     \label{fig:optimization_solution_func}
 \end{subfigure}
 \hfill
 \begin{subfigure}[b]{0.325\linewidth}
     \centering
     \includegraphics[width=\linewidth]{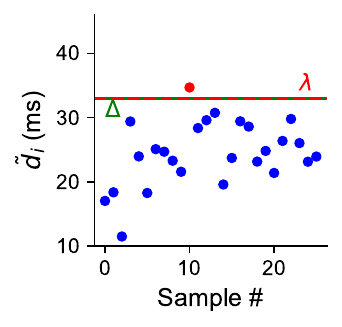}
     \caption{\centering \small Counterfactual samples}
     \label{fig:optimization_solution_after}
 \end{subfigure}

 \caption{\small 
 Finding target bitrate fraction ($\alpha$).
 Given a set of \fst samples (left), whose $\lambda^{th}$ percentile and outliers higher than $P = \Delta$ are shown in red, we evaluate find $\alpha$ that matches the counterfactual $\lambda^{th}$ percentile to $P$ (middle). We update the counterfactual values of \fst with $\alpha$ to have fewer outliers above $P$ (right).}
 \label{fig:optimization_solution}
 \vspace{-10pt}
\end{figure}

    \noindent \Fig{optimization_solution} shows an example of this counterfactual problem for $\lambda = 0.9$. \Fig{optimization_solution_before} shows the \fst samples and their $\lambda^{th}$ percentile (red line). The samples that are less than $P = \Delta$ are colored in blue, and those larger than $P$ (which would cause higher latency) are colored in red.
    \Fig{optimization_solution_func} shows the distribution of $d_i \frac{\text{\ccrate}}{tr_i}$ (called as normalized $d_i$) and its $\lambda^{th}$ percentile (denoted by $x$). Here, $\alpha$ is set to $\frac{P}{x}$. \Fig{optimization_solution_after} shows the counterfactual \fst values, $\Tilde{d_i}$, had $\alpha$ been used to encode them. The scaled-down $\Tilde{d_i}$ values reduce the number of outliers above $P$ (decreasing the latency).

\begin{figure}[t]
 \centering
 \vspace{-5 pt}
 \begin{subfigure}[b]{\linewidth}
     \centering
     \includegraphics[width=\linewidth]{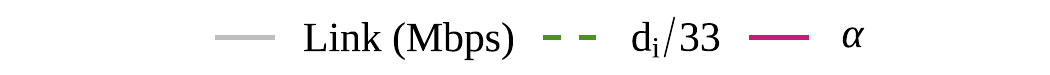}
     \label{fig:noise_alpha_legend}
 \end{subfigure}

 \begin{subfigure}[b]{\linewidth}
  \vspace{-15pt}
     \centering
     \includegraphics[width=\linewidth]{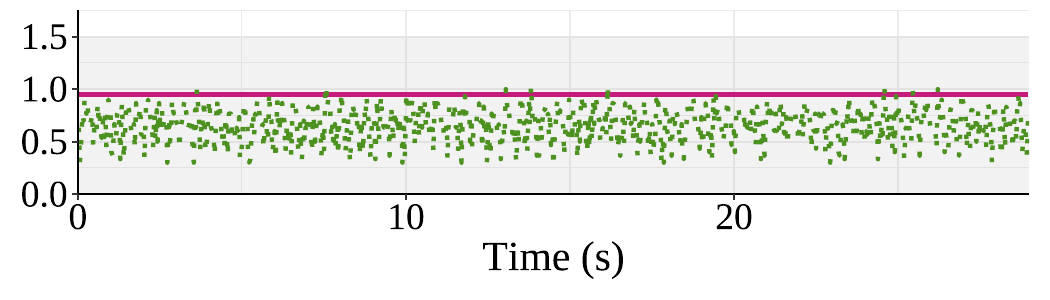}
     \caption{\small Fixed \SI{1.5}{Mbps} link with a low-motion video.}
     \vspace{-1 pt}
     \label{fig:fixed_alpha_link}
 \end{subfigure}

 \begin{subfigure}[b]{\linewidth}
     \centering
     \includegraphics[width=\linewidth]{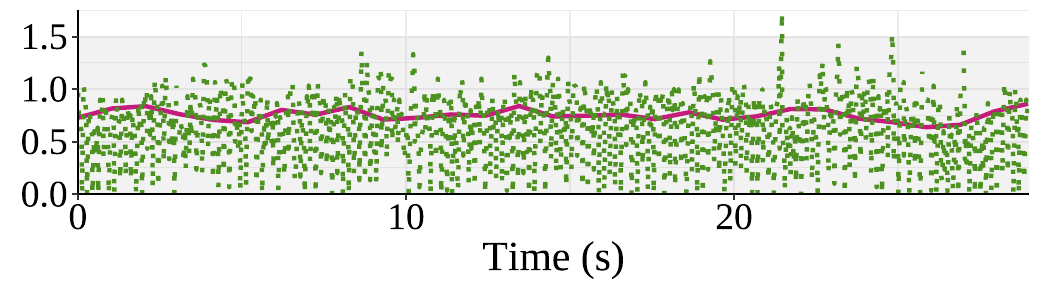}
     \caption{\small Fixed \SI{1.5}{Mbps} link with a high-motion video.}
     \vspace{-1 pt}
     \label{fig:noise_alpha_video}
 \end{subfigure}

 \begin{subfigure}[b]{\linewidth}
     \centering
     \includegraphics[width=\linewidth]{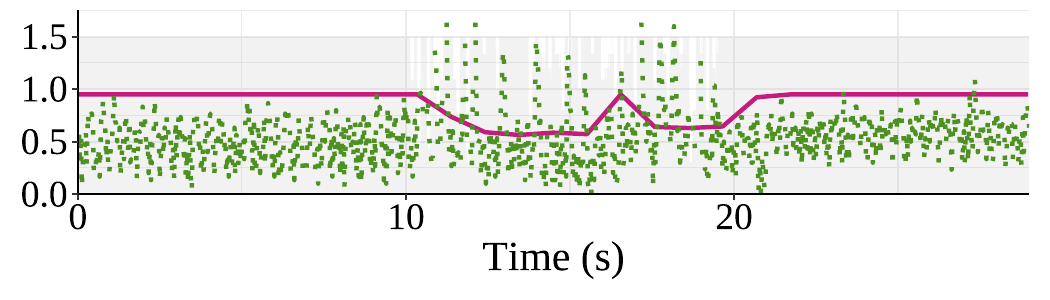}
     \caption{\small \SI{1.5}{Mbps} link with 10s of bandwidth fluctuation and a low-motion video.}
     \vspace{-5 pt}
     \label{fig:noise_alpha_link}
 \end{subfigure}

 \caption{\small $\alpha$'s response to link and video encoder variations. $\alpha$ picks lower values (lower target bitrate) when the link capacity or encoder output varies significantly to maintain good control over the \fsts. $d_i$ denotes the \fst in milliseconds.
 }
 \vspace{-15 pt}
 \label{fig:noise_alpha}
\end{figure}

\NewPara{How does $\alpha$ work?} $\alpha$ determines the headroom between the encoder's target bitrate ($\alpha \cdot \text{\ccrate}$) and the estimated network rate (\ccrate). \TheSystem automatically adjusts $\alpha$ to cope with variability caused by both changing video content and fluctuating network rates. To illustrate this, \Fig{noise_alpha} shows the behavior of \TheSystem in three scenarios. All three experiments use  $\lambda=0.9$ and $P = 33ms$. First, we run a fixed \SI{1.5}{Mbps} link and a low-motion video, where we repeatedly feed the encoder with a fixed $1280\times720$ frame to eliminate any encoder variance. \Fig{fixed_alpha_link} shows the values of $\alpha$ and the ratio of \fst to $\Delta = 33ms$. In steady conditions, $\alpha$ has a stable value of $\sim 1$ and the P90 of the \fsts is below \SI{33}{ms}. Next, we test \TheSystem on the same link with a high-motion video~\cite{dynamic_video} in \Fig{noise_alpha_video}. The encoded frame sizes and the frame service times are now much more variable. Consequently, $\alpha$ adapts to the variable encoder output by switching to a lower value to control the tail of \fsts.
Finally, we run \TheSystem on a \SI{1.5}{Mbps} link that experiences 10s of high variability in \Fig{noise_alpha_link}. Before the fluctuations start at 10s, \fst values are well below $\Delta$ with $\alpha \sim 1$. During the noisy period (10s--20s), when the \fsts increase, $\alpha$ decreases to improve the P90 of \fst and reduce video bitrate. When the link steadies after the 20s, $\alpha$ resets to its high value.

\bigskip
\section{Implementation}
\label{sec:impl}

We implemented our system on top of Google's WebRTC~\cite{webrtc}. 

\NewPara{Congestion Controller.} We implement two window-based delay-sensitive algorithms, Copa~\cite{arun2018copa} and RoCC~\cite{rocc}. We reused the logic from the original implementation of Copa~\cite{copa-implementation}. For RoCC, given $\text{RTT}_{min}$, the minimum observed RTT, RoCC sets the congestion window (\cwnd) to a small constant more than the number of bytes received in the last $(1 + \gamma) \text{RTT}_{min} $ interval. This maintain a network queueing delay of $\gamma \text{RTT}_{min}$ where $\gamma$ is the delay-sensitiveness parameter. We use Copa as the default unless otherwise mentioned.

\NewPara{Dummy Generator.} We repurpose the padding generator in WebRTC to generate dummy packets that are within the \cwnd and smaller than 200 bytes each.
Dummy packets are ACKed by the receiver, but their payload is discarded. The dummy packets have ``padding'' type in their packet header, and the receiver distinguishes them from the ``video'' packets.

\NewPara{Latency Safeguards.} 
The transport layer
sets the encoder target bitrate to zero to signal a \emph{pause} if the oldest packet's age in the pacer queue exceeds the pacer queue pause threshold ($\tau$). We reuse WebRTC's support for buffering the latest unencoded camera frame.

\NewPara{Encoder Rate Controller.}
\TheSystem records the target bitrate used for encoding each frame. It also measures \fst by recording the time that the first packet of a frame is at the front of the pacer and when the last packet of the frame leaves the pacer (this does not include the queuing time the frame spent in the pacer before frames ahead of it).  
\TheSystem picks the next $\alpha$ based on the recorded frame service times every time interval $T=1s$ using the algorithm in~\S\ref{sec:design tradeoff}. We further apply an EWMA smoothing over the values of $\alpha$ to ensure gradual changes in the encoder's target bitrate over time.

\section{Evaluation}
\label{sec:eval}

We conducted an evaluation of \TheSystem using a WebRTC-based setup on Mahimahi~\cite{netravali2015mahimahi} emulated links. Our experimental setup is detailed in \Sec{sec:eval_setup}, and we use this to benchmark against existing systems in \Sec{sec:eval_overall}. A thorough examination of \TheSystem's design features is presented in \Sec{sec:eval_system}. Additional evaluations focusing on less critical parameters ($\lambda$ and $\tau$) and results with alternative quality metrics are available in the appendix.

\begin{figure*}[t]
 \centering
 \begin{subfigure}[b]{\textwidth}
   \vspace{-5pt}
   \centering
   \includegraphics[width=0.5\linewidth]{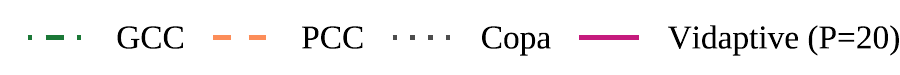}
   \vspace{-5 pt}
   \label{fig:cdf_legend}
 \end{subfigure}
 
 \begin{subfigure}[b]{0.34\linewidth}
   \centering
   \includegraphics[width=\linewidth]{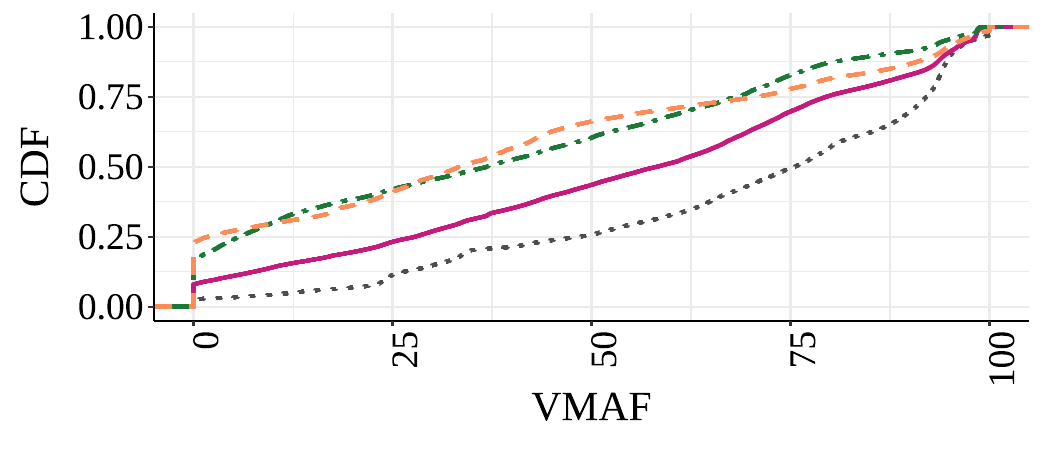}
   \caption{\small VMAF CDF}
   \label{fig:cdf_vmaf}
 \end{subfigure}%
 \begin{subfigure}[b]{0.34\linewidth}
   \centering
   \includegraphics[width=\linewidth]{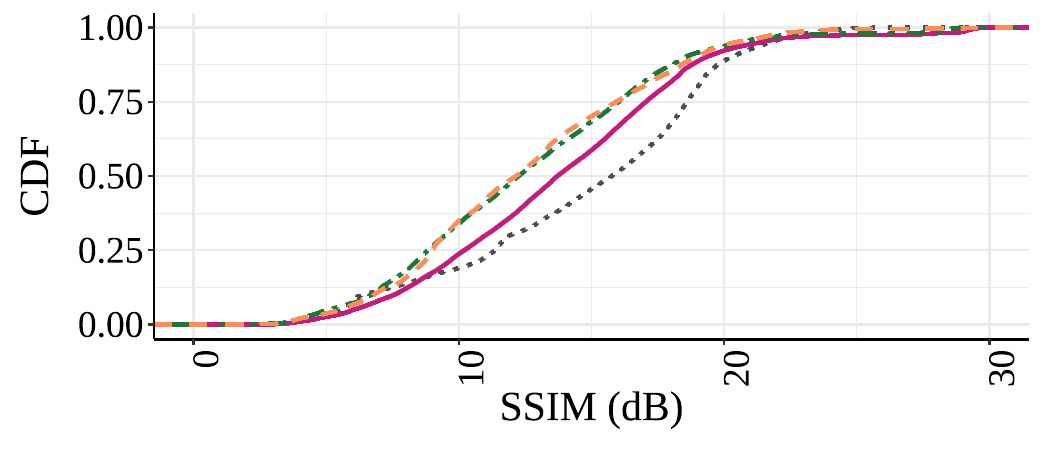}
   \caption{\small SSIM CDF}
   \label{fig:overall_cdf_ssim}
 \end{subfigure}%
 \begin{subfigure}[b]{0.34\linewidth}
   \centering
   \includegraphics[width=\linewidth]{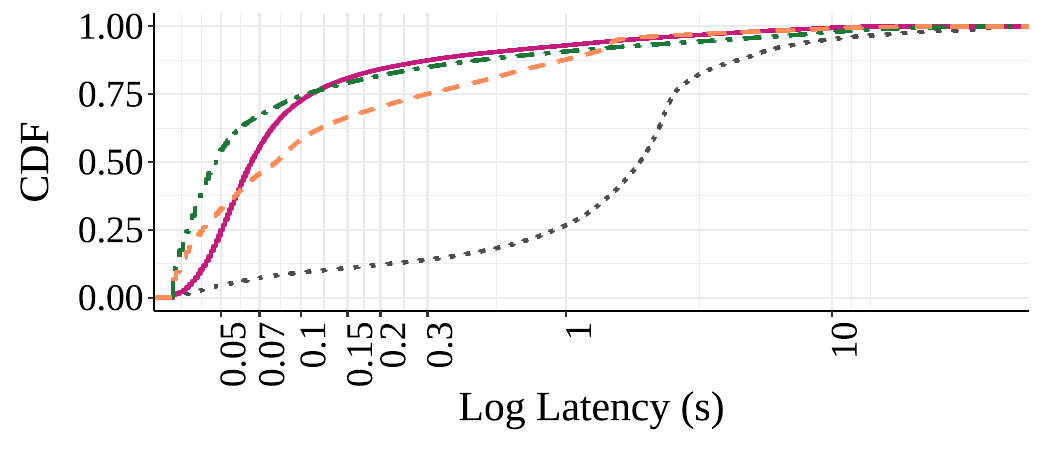}
   \caption{\small Frame Latency CDF}
   \label{fig:cdf_latency}
 \end{subfigure}

 \caption{\small CDF of frame VMAF and SSIM and latency across all the 1M frames. Compared to GCC and PCC, \TheSystem achieves higher VMAF and SSIM on all percentiles while getting better latency on higher percentiles. \TheSystem increases the median latency over GCC by a marginal amount of \SI{17}{ms}. Copa's higher VMAF scores come at a great latency cost, rendering it unsuitable for real-time video applications.}
 \label{fig:overall_cdfs}
 \vspace{-10pt}
\end{figure*}

\begin{figure}
    \centering
    \includegraphics[width=\linewidth]{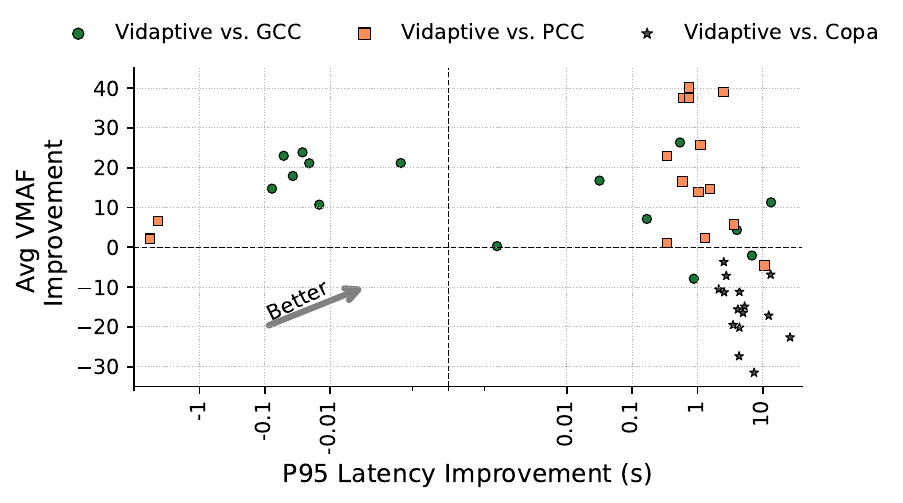}
    \vspace{-20 pt}
    \caption{\small Performance benefits of \TheSystem over GCC, PCC, and Copa. \TheSystem either improves latency or quality or both over all baselines. Against GCC, \TheSystem shows higher VMAF in 80\% of the traces, improves P95 latency on half the traces, and increases the P95 latency by less than \SI{100}{ms} on the other half. 
    \TheSystem also maintains significantly lower P95 latency (around 1 second) than Copa on all traces and is better than PCC in 86\% of traces. Each data point represents a unique trace, with the x-axis logarithmically scaled.}
    \label{fig:overall_vmaf_vs_latency}
    \vspace{-15pt}
\end{figure}

\subsection{Setup}
\label{sec:eval_setup}

\NewPara{Testbed.} We built a testbed, implemented in C++, on top of WebRTC~\cite{webrtc_git_commit} that enables a headless peer-to-peer video call between two endpoints.
The sender program emulates a camera by reading video frames from an input file
and the receiver records the received frames to an output file. To match video frames between the sender and the receiver for visual quality and latency measurements, a unique 2D barcode is placed on each frame~\cite{salsify}. 
We emulate different network conditions between the sender and receiver by placing the receiver behind a Mahimahi link shell. All experiments are run for \SI{2}{min} on a link with a one-way delay of \SI{25}{ms}. \TheSystem uses Copa~\cite{arun2018copa} as the default \cca, unless mentioned otherwise. 

\NewPara{Metrics.}
\label{eval_merics}
To assess the performance of \TheSystem, we focus on two key metrics: frame quality and frame latency. For frame quality, we use Video Multimethod Assessment Fusion (VMAF~\cite{vmaf}), comparing received frames to their original sources. Additionally, we report Structural Similarity Index Measure (SSIM~\cite{ssim}), and report the Peak Signal-to-Noise Ratio (PSNR~\cite{psnr}) in Appendix~\ref{sec:more_metrics}. Frame latency is gauged by the duration from \emph{frame read} at the sender to \emph{frame display} at the receiver, setting the display time for undelivered frames to the presentation time of the next frame displayed~\cite{salsify}. Moreover, we report the total video bitrate and \fr at the receiver.

\NewPara{Network Traces.} We evaluate each scheme on a set of 15 cellular traces bundled with Mahimahi~\cite{netravali2015mahimahi} and also use synthetic traces to illustrate the convergence behavior in Appendix~\ref{eval_convergence}. For the cellular traces, we use a large bottleneck buffer size that does not incur packet loss. We also evaluate each scheme on three Pantheon’s calibrated network emulation traces~\cite{pantheon}. These traces were tuned to mimic real Internet paths. They have limited buffers (and thus incur congestion-induced loss) as well as stochastic packet loss.

\NewPara{Videos.}
We utilize a dataset of 20 different 1080p videos,
each playing at 30 frames per second. The dataset is curated from YouTube and spans different levels of motion intensity across sports, cooking, video conferencing, nature documentaries, live performances, and screen sharing. \Tab{dataset_info} in describes the dataset in detail.
Unless specified otherwise, all experiments are conducted on the complete set of videos from this dataset (\Tab{dataset_info}). Audio is disabled throughout the experiments.

\NewPara{Baselines.} In our evaluation, \TheSystem is benchmarked against several established congestion control algorithms to demonstrate their respective capabilities in handling real-time video scenarios over variable links. First, we consider the Google Congestion Control (GCC) algorithm, the standard for WebRTC transport~\cite{gcc}. The GCC protocol has changed since publication~\cite{gcc}; we used the latest implementation in the M108~\cite{webrtc_git_commit} version of WebRTC. Additionally, we compare \TheSystem  to Performance-oriented Congestion Control (PCC), another rate-based approach~\cite{pcc}. We further include an analysis against Copa, a state-of-the-art delay-based congestion control algorithm~\cite{arun2018copa}. All algorithms are implemented within the same WebRTC framework.

\subsection{Overall Comparison}
\label{sec:eval_overall}

\NewPara{Cellular Traces.} In \Fig{overall_cdfs}, we present a comparative analysis of \TheSystem's performance against GCC, PCC, and Copa using WebRTC on the cellular traces bundled in Mahimahi~\cite{mahimahi_traces}. \Fig{cdf_vmaf} shows the cumulative distribution functions (CDFs) for VMAF, SSIM, and frame latency, encompassing all frames and traces. Compared to GCC and PCC, \TheSystem achieves a higher VMAF and SSIM across all percentiles by utilizing the link more effectively. It improves the average VMAF to 53.5 vs. $\sim$39 for GCC and PCC, and improves the median (P50) VMAF to 58.2, up from 36.8 (GCC) and 33.4 (PCC). Regarding SSIM, \TheSystem raises the average to 14 from 12 and boosts the P50 SSIM to 13.8 from 12, vs. GCC and PCC. Copa achieves higher VMAF and SSIM scores than \TheSystem. However, it has a significantly higher latency (e.g., an average latency of 2.9 seconds), rendering it impractical for most real-time video applications. The reason is that, with video traffic, Copa is application-limited most of the time. Copa's congestion control algorithm does not work properly in this regime. Since it does not detect any queueing delay, it simply increases its congestion window size to a maximum cap. This causes excessive queueing when network conditions deteriorate. \TheSystem underlying CCA is also Copa, but it is able to control delays effectively since the source behaves like a backlogged flow. 

\TheSystem exhibits a slight \SI{17}{ms} higher median latency than GCC (\SI{65}{ms} vs. \SI{48}{ms}), yet it significantly improves tail frame latency, particularly above the $75^\text{th}$ percentile. These upper percentiles correspond to conditions with high link rate variability and outages, where \TheSystem can respond faster to such fluctuations. For instance, at the $95^\text{th}$ percentile, \TheSystem cuts down frame latency by \SI{2246}{ms} relative to GCC, reducing it from \SI{3941}{ms} to \SI{1695}{ms}. It also betters the average latency by \SI{351}{ms} (\SI{383}{ms} vs. GCC's \SI{734}{ms}). Compared to PCC, \TheSystem lowers both the average latency (\SI{383}{ms} versus \SI{521}{ms}) and the median latency (\SI{65}{ms} vs. \SI{81}{ms}).

\Fig{overall_vmaf_vs_latency} shows the P95 latency and average VMAF improvement of \TheSystem across all the videos for each trace. Each point represents a different trace. \TheSystem either improves VMAF or latency (or both) against all baselines. 
Compared to GCC, \TheSystem achieves higher VMAF on 80\% of the traces, improves P95 latency on half the traces, and increases the P95 latency by less than \SI{100}{ms} on the other half.
\TheSystem reduces P95 latency significantly (by approximately 1 second) in all traces against Copa and in 86\% of traces (13 out of 15) against PCC. Additionally, \TheSystem shows higher VMAF in 14 out of 15 traces than PCC.

\begin{figure}[t]
 \centering
 \begin{subfigure}[b]{\linewidth}
   \vspace{-5 pt}
     \centering
     \includegraphics[width=\linewidth]{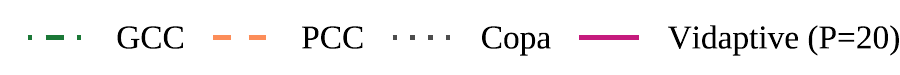}
      \vspace{-10pt}
     \label{fig:cdf_legend_pantheon}
 \end{subfigure}

 \begin{subfigure}[b]{0.47\linewidth}
     \centering
     \includegraphics[width=\linewidth]{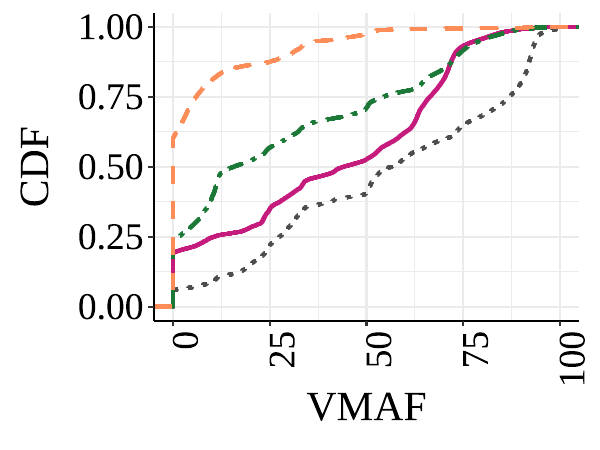}
     \caption{\small VMAF CDF }
     \label{fig:cdf_vmaf_pantheon}
 \end{subfigure}
 \hfill
 \begin{subfigure}[b]{0.47\linewidth}
  \vspace{-5pt}
     \centering
     \includegraphics[width=\linewidth]{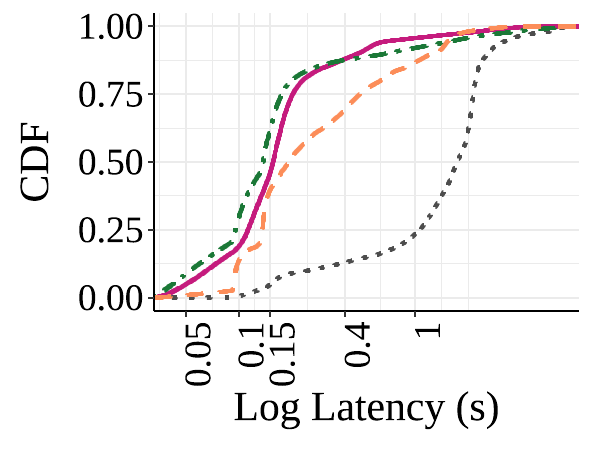}
     \caption{\small Latency CDF}
     \label{fig:cdf_latency_pantheon}
 \end{subfigure}

 \caption{\small CDF of VMAF and latency for all the frames in Pantheon traces. \TheSystem improves the quality over PCC in all the percentiles. Compared to GCC, \TheSystem has better quality in low-medium percentiles and similar quality in high percentiles. Copa has higher quality at the cost of significant latency. \TheSystem improves the latency in all percentiles vs. Copa and PCC. Compared to GCC, \TheSystem improves $95^\text{th}$ by $\sim$ \SI{1}{s} while increasing the median latency \SI{19}{ms} vs. GCC.}
 \label{fig:pantheon_cdfs}
 \vspace{-15 pt}
\end{figure}

\NewPara{Pantheon's Calibrated Emulators.} 
We extend our tests to three emulated network paths from Pantheon~\cite{pantheon} that are calibrated to approximate various real Internet paths. Unlike the cellular traces, these links have constant capacity with a Poisson delivery process but experience loss and have shallow queues with tail drop. \Fig{pantheon_cdfs} displays the performance comparison of \TheSystem along with GCC, PCC, and Copa on three of such traces: ``Nepal to AWS India (Wi-Fi)''~\cite{pantheon_nepal_to_india}, ``Mexico to AWS California (Cellular)''~\cite{pantheon_mexico_to_california}, and ``AWS Brazil to Colombia (Cellular)''~\cite{pantheon_brazil_to_colombia} on five videos. \TheSystem achieves a median VMAF of 44 over GCC's 15 and an average VMAF of 40 over GCC's 29. Both GCC and \TheSystem achieve a P95 VMAF of 79. Compared to PCC, \TheSystem achieves better VMAF on all the percentiles.

\Fig{cdf_latency_pantheon} shows that \TheSystem achieves \SI{933}{ms} lower P95 latency (\SI{1739}{ms} $\rightarrow$ \SI{806}{ms}), and \SI{89}{ms} (\SI{362}{ms} $\rightarrow$ \SI{273}{ms}) lower average latency than GCC, while increasing the P50 latency by \SI{19}{ms} (\SI{137}{ms} $\rightarrow$ \SI{156}{ms}). This slight increase in the median latency is consistent with our observations in the cellular traces and comes from the small queue that \TheSystem's \cca (Copa) keeps in the network to track changes in available bandwidth. Compared to PCC and Copa, \TheSystem achieves a lower latency on all the percentiles.
\subsection{Understanding \TheSystem's Design}
\label{sec:eval_system}
\begin{figure*}[t]
 \centering
 \begin{subfigure}[b]{0.7\linewidth}
     \centering
     \includegraphics[width=\linewidth]{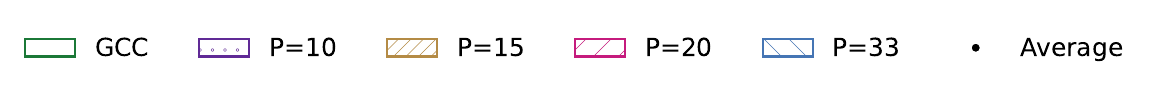}
     \vspace{-10 pt}
     \label{fig:Pacer_Serve_time_legend}
 \end{subfigure}

 \begin{subfigure}[b]{0.19\linewidth}
     \centering
     \includegraphics[width=\linewidth]{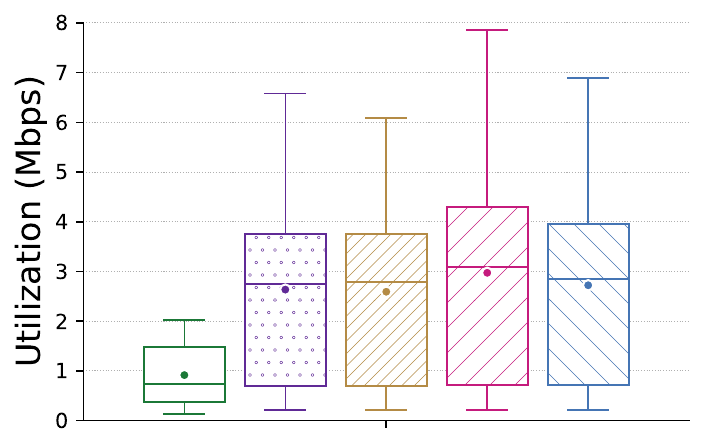}
     \caption{\small Link Utilization}
     \label{fig:Pacer_Serve_time_summary_utilization}
 \end{subfigure}
 \begin{subfigure}[b]{0.19\linewidth}
     \centering
     \includegraphics[width=\linewidth]{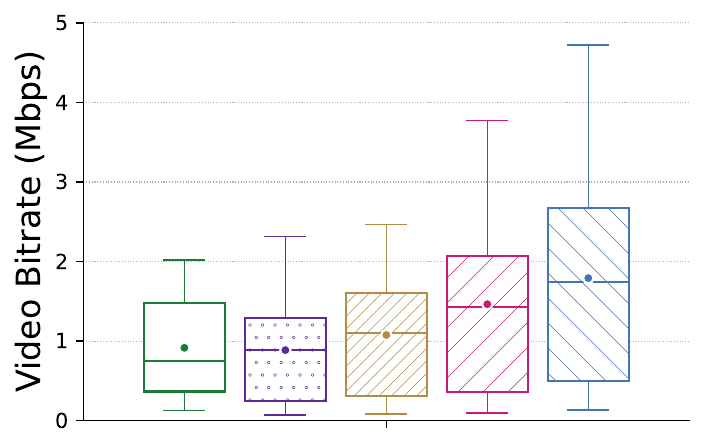}
     \caption{\small Video Bitrate}
     \label{fig:Pacer_Serve_time_summary_video}
 \end{subfigure}
 \begin{subfigure}[b]{0.19\linewidth}
     \centering
     \includegraphics[width=\linewidth]{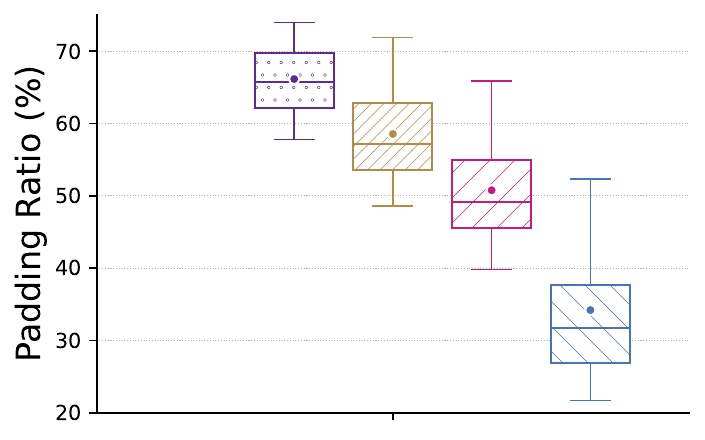}
     \caption{\small Padding Ratio}
     \label{fig:Pacer_Serve_time_summary_padding_ratio}
 \end{subfigure}
 \begin{subfigure}[b]{0.19\linewidth}
     \centering
     \includegraphics[width=\linewidth]{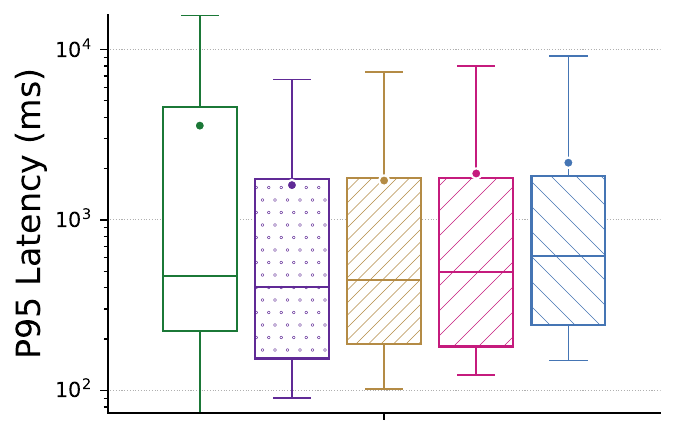}
     \caption{\small Log P95 Latency}
     \label{fig:Pacer_Serve_time_summary_latency}
 \end{subfigure}
 \begin{subfigure}[b]{0.19\linewidth}
     \centering
     \includegraphics[width=\linewidth]{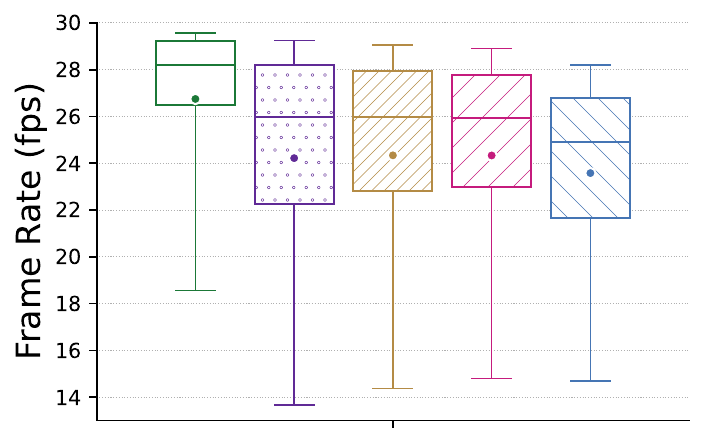}
     \caption{\small Frame Rate}
     \label{fig:Pacer_Serve_time_summary_frame_rate}
 \end{subfigure}
\vspace{-5pt}
 \caption{\small Performance comparison of GCC and \TheSystem with different values of target \fst $P$. The whiskers indicate $5^\text{th}$ and $95^\text{th}$ percentile (P5 and P95) values, the interquartile range covers the $25^\text{th}$ to $75^\text{th}$ percentile (P25--P75), the horizontal line marks the median (P50), and the bullet symbol ($\bullet$) represents the average value.}
 \label{fig:Pacer_Serve_time_overview}
\end{figure*}

\begin{figure}[t]
\centering
    \includegraphics[width=0.9\linewidth]{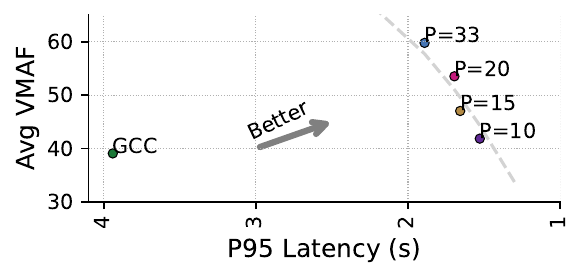}
    \vspace{-10pt}
    \caption{\small Average VMAF vs P95 Latency over all the frames. \TheSystem can tradeoff latency and quality by parameter $P$ and it offers a better tradeoff than the other schemes.}
\label{fig:Pacer_Serve_time_frontier}
\end{figure}

\begin{figure}[t]
    \centering
    \includegraphics[width=\linewidth]{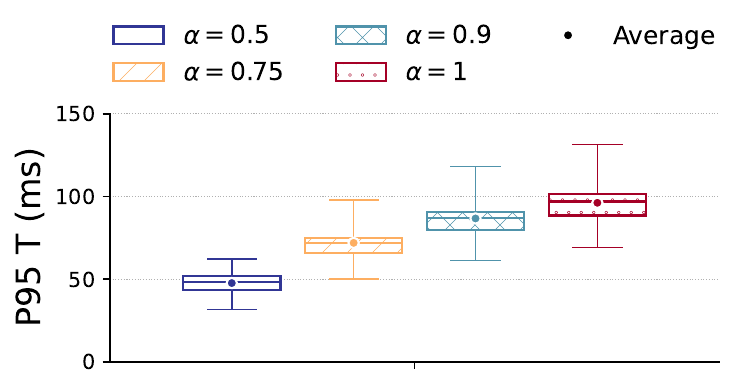}
    \vspace{-10pt}
    \caption{\small
    Illustrating the relationship between the target bitrate factor $\alpha$ and the distribution of the $95^\text{th}$ percentile (P95) values of transmission time $T$. As $\alpha$ increases, indicating a higher fraction of the link capacity utilized for video, the P95 values of $T$ shift towards higher latency numbers. This demonstrates the trade-off between maximizing video bitrate and controlling latency, especially under conditions of variable network bandwidth.}
    \label{fig:P_bar}
    \vspace{-10pt}
\end{figure}

\NewPara{Frame Serve Time Threshold ($P$).}
In our analysis of \TheSystem, we first focus on the impact of varying the target \fst ($P$), which determines the tradeoff preference of latency and quality (\S\ref{sec:design tradeoff}).
\Fig{Pacer_Serve_time_overview} illustrates the distribution of metrics for all the traces and all videos (15 $\times$ 20 settings). We use Mahimahi's cellular traces for the rest of the paper.

Notably, due to the presence of dummy traffic---which decouples link utilization from the video application's decision-making---changing $P$ does not affect overall link utilization. This results in a consistently higher utilization than GCC, as depicted in~\Fig{Pacer_Serve_time_summary_utilization}.

As highlighted in \S\ref{sec:design tradeoff}, an increase in $P$ leads \TheSystem to choose higher video bitrates (\Fig{Pacer_Serve_time_summary_video}). This reduces the proportion of dummy traffic (\Fig{Pacer_Serve_time_summary_padding_ratio}). As $P$ rises, \TheSystem prioritizes video bitrate over latency, increasing P95 latency (\Fig{Pacer_Serve_time_summary_latency}). However, \TheSystem still maintains a lower P95 latency than GCC.

The \fr remains relatively stable with increasing $P$ values, but as $P$ approaches \SI{33}{ms}, the frame skipping mechanism is more likely to be activated, causing a slight decrease in \fr for $P=33$ (\Fig{Pacer_Serve_time_summary_frame_rate}). \Fig{Pacer_Serve_time_frontier} presents the average VMAF vs. P95 of latency of GCC and \TheSystem using different values of $P$ over all the frames. This illustrates that \TheSystem achieves a better performance frontier than GCC.

An interesting question is: \emph{why is it not possible to increase the video bitrate (send less dummy traffic) without affecting the tail latency?} The answer lies in the nature of highly fluctuating links, like those in our evaluation traces. Raising the target bitrate too close to the link capacity risks latency spikes during sudden capacity drops. For instance, a frame encoded for a \SI{2}{Mbps} link taking \SI{33}{ms} to transmit, will need \SI{132}{ms} on a suddenly reduced \SI{500}{Kbps} link, leading to an unavoidable increase in latency. As also noted in prior work~\cite{salsify}, this issue stems from the encoding process of traditional video codecs: once a frame is encoded, it cannot be undone. The sender is then left with two choices: either send the oversized frame, causing self-inflicted delay and loss, or drop it and reinitialize the stream with a keyframe, which typically results in an even larger frame size.

To illustrate this point quantitatively, imagine a scenario where the transport layer has precise knowledge of the link rate, $C(t)$, at every moment $t$.
Coupled with this, assume the existence of an ``ideal'' encoder capable of generating ``ideal'' frames with precise sizes.\footnote{This ideal encoder also encodes frames instantaneously since the link rate estimate could change \emph{during} the encoding process.
}
When this encoder is given a target bitrate of $\alpha C(t)$, it produces a frame of size $\alpha C(t)\Delta$, with $\Delta$ representing the frame interval. Let $T(\alpha, t)$ denote the transmission time for an ideal frame produced at time $t$. Assuming no other data is being sent at the time of transmission, $T(\alpha,t)$ can be derived as follows:
\begin{equation}
\vspace{-5 pt}
\int_{t}^{t + T(\alpha, t)} C(\tau) \, d\tau = \overbracket{{\alpha C(t)}  \cdot \Delta}^\text{frame size}
\label{eqn:find_T}
\end{equation}

We studied the values of $T(\alpha, t)$ for $\alpha = 0.5, 0.75, 0.9, 1.0$ across all the traces to understand their characteristics. \Fig{P_bar} shows the distribution of P95 values of $T$ across all the traces for different values of $\alpha$. As $\alpha$ increases, indicating a higher fraction of the link's capacity being utilized, the P95 values of $T$ also rise. This demonstrates that, despite an ideal transport and encoder setup, $T$---and consequently, \fst---are lower bounded by the inherent variability of the link, and the tail latency will increase as $\alpha$ increases.

To manage tail latency effectively, \TheSystem  therefore has to pick lower values for $\alpha$ on some traces. However, this decision implies that it will also need to send more dummy traffic to compensate for the resulting under-utilization of video traffic on such links. Sending this dummy traffic allows the \cca to track the network capacity accurately, but it has minimal effect on frame latency since the dummy packets only get sent when there isn't a video frame available (\S\ref{sec:design transport}).

\begin{figure}[t]
 \centering
 \begin{subfigure}[b]{\linewidth}
     \centering
     \includegraphics[width=\linewidth]{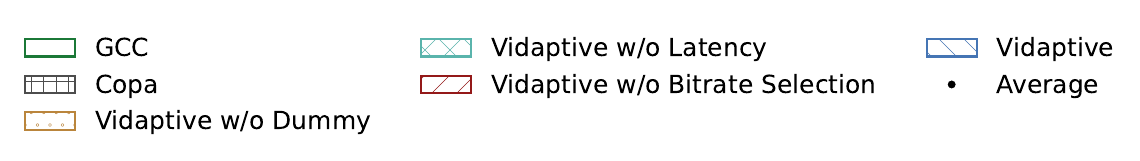}
     \label{fig:ablation_legend}
 \vspace{-10pt}
 \end{subfigure}
 \begin{subfigure}[b]{0.47\linewidth}
     \centering
     \includegraphics[width=\linewidth]{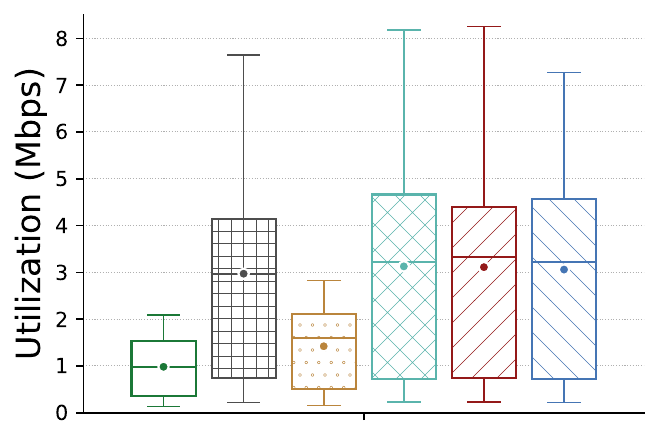}
     \caption{\small Link Utilization}
     \label{fig:ablation_summary_util}
 \end{subfigure}
 \quad
 \begin{subfigure}[b]{0.47\linewidth}
     \centering
     \includegraphics[width=\linewidth]{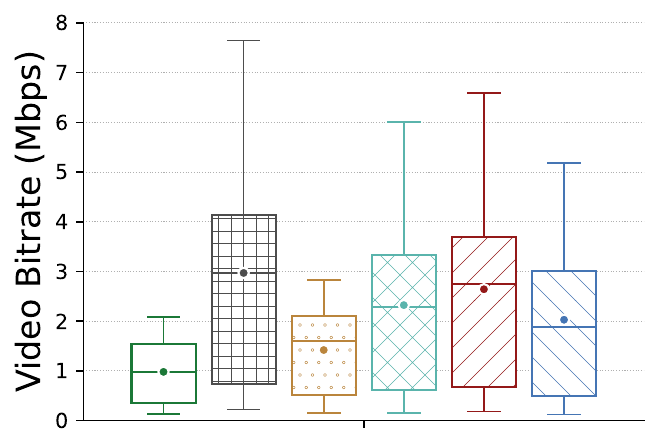}
     \caption{\small  Video Bitrate}
     \label{fig:ablation_summary_video}
 \end{subfigure}

 \begin{subfigure}[b]{0.47\linewidth}
     \centering
     \includegraphics[width=\linewidth]{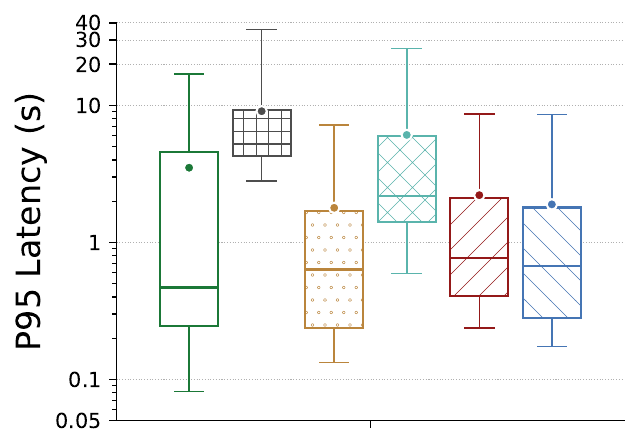}
     \caption{\small Log of P95 Latency}
     \label{fig:ablation_summary_latency}
 \end{subfigure}
 \quad
 \begin{subfigure}[b]{0.47\linewidth}
     \centering
     \includegraphics[width=\linewidth]{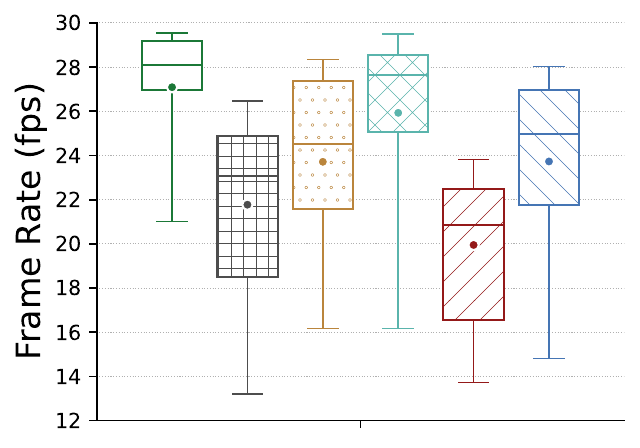}
     \caption{\small Frame Rate}
     \label{fig:ablation_summary_frame_rate}
 \end{subfigure}

 \caption{\small 
 \small Performance comparison of GCC and different \TheSystem components. ``Copa'' improves video bitrate and utilization but hurts frame latency. Dummy traffic improves video bitrate and utilization. ``\TheSystem w/o Latency'' shows that  latency knobs reduce the latency by \emph{seconds}. \TheSystem has higher \fr and lower latency by using encoder bitrate selection than without it. Since schemes with Copa do not send frames in outages, they have lower \fr than GCC. The whiskers are P5 and P95, the interquartile range shows P25, P50, and P75.
 }
 \label{fig:ablation_overview}
 \vspace{-10pt}
\end{figure}

\NewPara{Ablation Study.}
To understand the impact of different components in \TheSystem's design,
we evaluate the benefits of changing the congestion control and adding dummy traffic at the transport layer (\Sec{sec:design transport}), enabling the latency safeguards (\Sec{sec:design latency}), and running the encoder bitrate selection approach described in \Sec{sec:design tradeoff}.
\Fig{ablation_overview} shows the distribution of metrics over all the traces and five videos for different system variations. $P$ is selected as \SI{33}{ms} for the rest of experiments.

In ``Copa,'' we replace GCC with a window-based congestion control algorithm but keep the remaining modules unchanged. Copa is more aggressive than GCC in allocating bandwidth, improving the average link utilization and video bitrate by almost $3\times$. However, this aggressiveness causes an average P95 latency of \emph{9 seconds}. The \fr for all schemes using Copa as CCA reduces because Copa's window-based mechanism, unlike GCC, stops frames from going out when congested.

In ``\TheSystem w/o Dummy,'' we removed dummy traffic (\Sec{sec:design transport}) from \TheSystem. \Fig{ablation_summary_video} shows that removing the dummy traffic reduces the average \emph{video} bitrate from \SI{2}{Mbps} to \SI{1.4}{Mbps} and its P95 from \SI{5.1}{Mbps} to \SI{2.8}{Mbps}. However, the average P95 latency and the \fr do not change significantly compared to \TheSystem: the average P95 latency decreases by only 5.6\% when dummy traffic is removed (from \SI{1894}{ms} to \SI{1788}{ms}). This slight increase results from \cca keeping a small queue in the network when it uses dummy traffic that causes a bigger spike (and P95 latency) when the network degrades abruptly. \textbf{This indicates that the dummy traffic has a negligible impact on latency while improving the video bitrate by discovering bandwidth faster.} 

In ``\TheSystem w/o Latency,'' we disable the latency safeguards on top of \TheSystem. Disabling latency knobs hurts the P95 latency (\Fig{ablation_summary_latency}) and increases it compared to \TheSystem (\SI{1894}{ms} vs. \SI{6087}{ms} on average and \SI{8.6}{s} vs. \SI{26}{s} in P95). Since the latency safeguards pause encoding of frames, the overall \fr and video bitrate increases without them.

In ``\TheSystem w/o Bitrate Selection,'' we disable the encoder bitrate selection of \TheSystem. Without this mechanism, the CCA's link estimate is directly given to the encoder, which results in a higher video bitrate and latency in highly variable links. When \ccrate is fed directly to the encoder, the encoder attempts to produce frames that take \SI{33}{ms} to be sent at the current link rate. However, if the link rate drops, such an encoded frame takes more than \SI{33}{ms} to be transmitted. This, in turn, triggers the latency safeguards that drop the next frames to prevent the latency from increasing. As a result, the \fr in ``\TheSystem w/o Bitrate Selection'' is lower than \TheSystem.

Finally, in ``\TheSystem'', the system aims to find the right target video bitrate for the encoder as described in \Sec{sec:design tradeoff}. Because this system balances frame latency and frame quality, the video bitrate is slightly lower than ``\TheSystem w/o Latency'' and ``Copa'' but achieves better latency.
The utilization is comparable across all schemes with dummy traffic since it pads any encoder output to match the link rate.

\begin{figure}[t]
 \centering
 \begin{subfigure}[b]{\linewidth}
     \centering
     \includegraphics[width=1\linewidth]{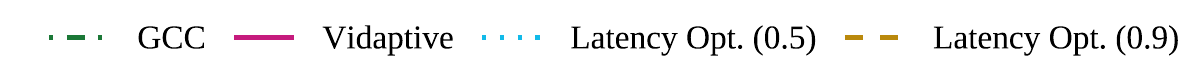}
      \vspace{-10pt}
     \label{fig:cdf_legend_alternative}
 \end{subfigure}

 \begin{subfigure}[b]{0.47\linewidth}
     \centering
     \includegraphics[width=\linewidth]{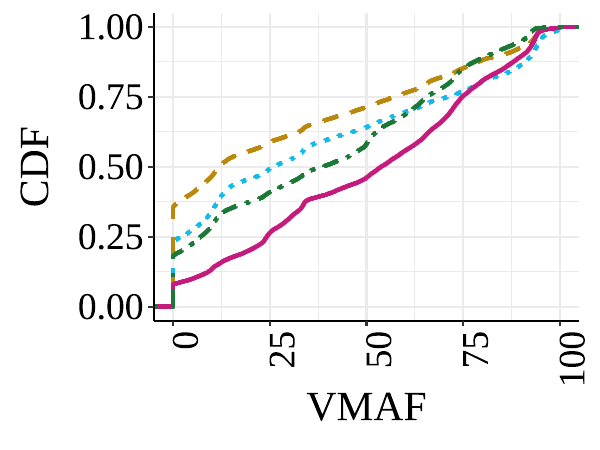}
     \caption{\small VMAF CDF }
     \label{fig:cdf_vmaf_alternative}
 \end{subfigure}
 \hfill
 \begin{subfigure}[b]{0.47\linewidth}
  \vspace{-5pt}
     \centering
     \includegraphics[width=\linewidth]{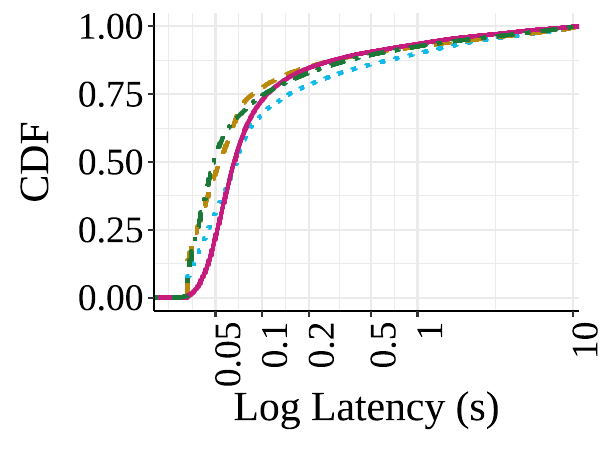}
     \caption{\small Latency CDF}
     \label{fig:cdf_latency_alternative}
 \end{subfigure}

 \caption{\small CDF of VMAF and latency for all the frames. The alternative latency optimizer scheme performs worse than \TheSystem.}
 \label{fig:alternative_cdfs}
 \vspace{-10 pt}
\end{figure}

Another system variation is disabling the dummy traffic and removing the pacer. Instead of controlling the \fst, this system runs the bitrate selection scheme described in \S\ref{sec:design tradeoff} on the measured one-way frame latencies instead of \fst. This ``Latency Optimizer'' scheme picks the next target bitrate based on the measured frame latencies and associated target bitrate.  \Fig{cdf_latency_alternative} shows the comparison of the ``Latency Optimizer'' scheme and \TheSystem along with GCC for $\lambda = 0.5, 0.9$. For $\lambda = 0.9$, ``Latency Optimizer'' opts to control P90 of measured latencies, and although the latency values are low, the quality drops significantly even compared to GCC. For $\lambda = 0.5$, ``Latency Optimizer'' targets a median latency of \SI{33}{ms}, but the tail latency increases compared to \TheSystem and GCC without meaningful gains in quality. The reason why this mechanism fails to work is that the bitrate selection mechanism alone is slow to adapt to high variations in the link as it looks at the latency measurements that come every \SI{33}{ms}, making it sluggish to increase the video bitrate (and thus, quality) when the link rate increases and decrease the bitrate in the face of sharp fluctuations.

\begin{figure}[t]
 \centering
 \begin{subfigure}[b]{\linewidth}
   \vspace{-5 pt}
     \centering
     \includegraphics[width=\linewidth]{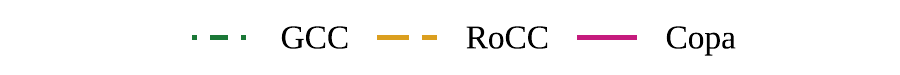}
      \vspace{-10pt}
     \label{fig:cdf_legend_rocc}
 \end{subfigure}

 \begin{subfigure}[b]{0.47\linewidth}
     \centering
     \includegraphics[width=\linewidth]{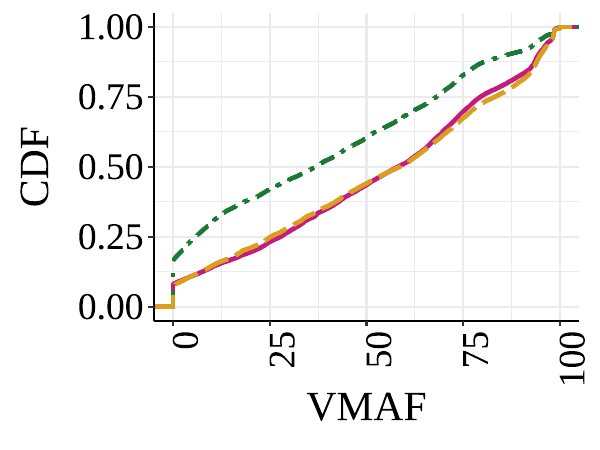}
     \vspace{-5 pt}
     \caption{\small VMAF CDF }
     \label{fig:cdf_vmaf_rocc}
 \end{subfigure}
 \hfill
 \begin{subfigure}[b]{0.47\linewidth}
  \vspace{-5pt}
     \centering
     \includegraphics[width=\linewidth]{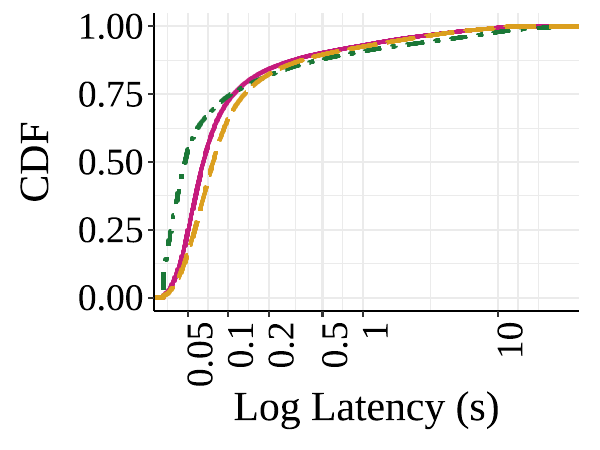}
     \caption{\small Latency CDF}
     \label{fig:cdf_latency_rocc}
 \end{subfigure}

 \caption{\small CDF of frame VMAF and latency across all the 1M 
 frames. Compared to GCC, \TheSystem(RoCC) achieves higher VMAF on all percentiles while getting lower latency on higher percentiles (worse latencies are better). \TheSystem(RoCC) has a similar performance to \TheSystem(Copa).}
 \label{fig:rocc_cdfs}
 \vspace{-10 pt}
\end{figure}

\NewPara{Using a Different Congestion Controller.} To show that \TheSystem can work with other delay-sensitive window-based \cca, we replaced Copa with RoCC~\cite{rocc}. 
\Fig{cdf_vmaf_rocc} shows the VMAF and \Fig{cdf_latency_rocc} shows the latency CDFs of \TheSystem(RoCC) compared to GCC. \TheSystem(RoCC) follows similar trends as \TheSystem(Copa) and improves the quality in all percentiles compared to GCC. Compared to GCC,  \TheSystem(RoCC) increases the median latency from \SI{48}{ms} to \SI{77}{ms} while reducing the P95 of latency from \SI{3.9}{s} to \SI{1.9}{s}. Compared to GCC, \TheSystem(RoCC) increases the median VMAF from 36 to 56 and the average VMAF from 39 to 53. Copa overall has a lower latency than RoCC, so we choose it as the default \cca in \TheSystem.

\begin{figure}[t]
 \centering
 \begin{subfigure}[b]{\linewidth}
   \vspace{-5 pt}
     \centering
     \includegraphics[width=\linewidth]{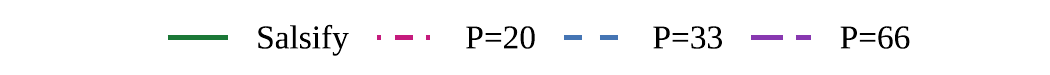}
      \vspace{-10pt}
     \label{fig:cdf_legend_salsify}
 \end{subfigure}

 \begin{subfigure}[b]{0.47\linewidth}
     \centering
     \includegraphics[width=\linewidth]{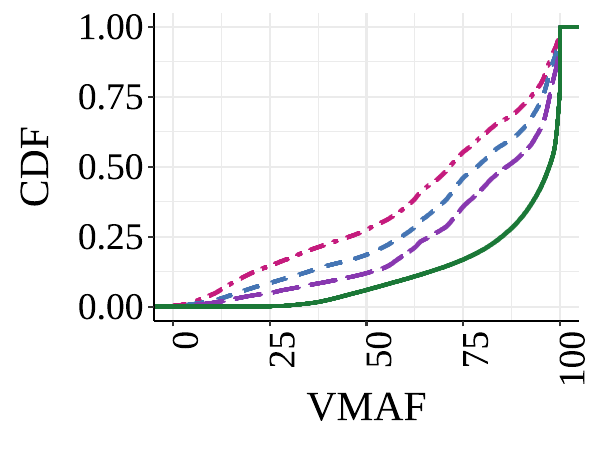}
     \caption{\small VMAF CDF }
     \label{fig:cdf_vmaf_salsify}
 \end{subfigure}
 \hfill
 \begin{subfigure}[b]{0.47\linewidth}
  \vspace{-5pt}
     \centering
     \includegraphics[width=\linewidth]{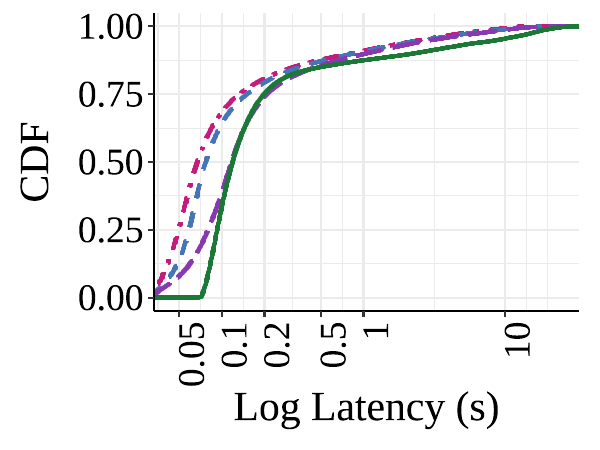}
     \caption{\small Latency CDF}
     \label{fig:cdf_latency_salsify}
 \end{subfigure}

 \caption{\small CDF of VMAF and latency for all the frames for Salsify and \TheSystem. Although Salsify provides better quality, it also experiences significant latency compared to \TheSystem.}
 \label{fig:salsify_cdfs}
 \vspace{-15 pt}
\end{figure}

\begin{figure}
    \centering
    \includegraphics[width=\linewidth]{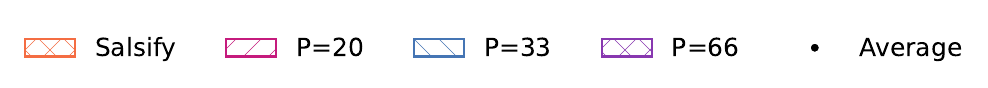}
    \label{fig:salsify_metrics_legend}
    \vspace{-20 pt}

\begin{subfigure}[b]{0.325\linewidth}
    \vspace{-5 pt}
     \centering
     \includegraphics[width=\linewidth]{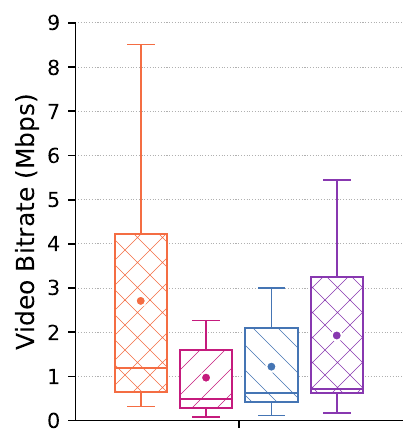}
     \caption{\centering  \small Video Bitrate}
     \label{fig:salsify_metrics_video}
 \end{subfigure}
 \hfill
 \begin{subfigure}[b]{0.325\linewidth}
     \centering
     \includegraphics[width=\linewidth]{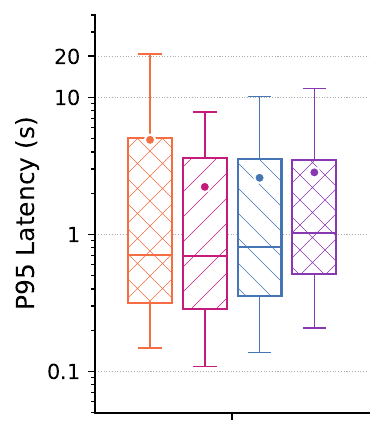}
     \caption{\centering \small Log P95 Latency
     }
     \label{fig:salsify_metrics_latency}
 \end{subfigure}
 \hfill
 \begin{subfigure}[b]{0.325\linewidth}
     \centering
     \includegraphics[width=\linewidth]{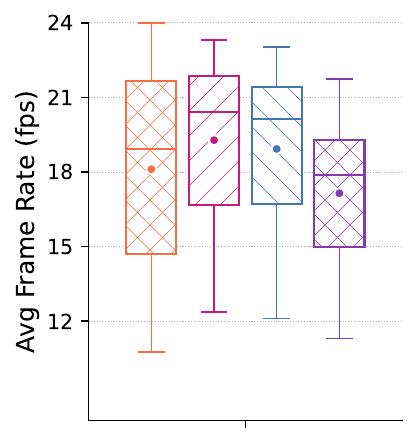}
     \caption{\centering Frame Rate}
     \label{fig:salsify_metrics_frame_rate}
\end{subfigure}
     
    \caption{\small Performance comparison of Salsify and \TheSystem on all the traces and videos. Salsify is experiencing much higher P95 latencies and lower \fr.
    The whiskers are P5 and P95, the interquartile range shows P25, P50, and P75.}
    \label{fig:salsify_metrics}
\end{figure}

\NewPara{Comparison with Salsify.} Salsify's design takes advantage of a specialized video codec that allows it to discard already-encoded frames when necessary. Since the peak performance of this encoder was 24 \fps at 720p, we focus specifically on videos of this size and frame rate. \Fig{salsify_cdfs} shows the CDF of VMAF and latency across all the frames. Although Salsify exhibits a higher quality (\Fig{cdf_vmaf_salsify}), it also experiences higher latency compared to \TheSystem (\Fig{cdf_latency_salsify}). The median latency for Salsify is \SI{119}{ms}, and the average latency is \SI{1211}{ms} across all the frames~\footnote{We are evaluating Salsify on a larger set of traces than Salsify's paper.}. However, \TheSystem has a median latency of \SI{67}{ms}, \SI{77}{ms}, \SI{117}{ms} and average latency of \SI{468}{ms}, \SI{535}{ms}, and \SI{622}{ms} for $P= 20, 33, 66$ respectively. \Fig{salsify_metrics_latency} shows the P95 latency distribution across all the videos and traces. Salsify experiences \emph{seconds} higher P95 latency than \TheSystem.

\Fig{salsify_metrics_frame_rate} shows the \fr for Salsify and \TheSystem. Salsify's primary means of latency control is dropping frames, which manifests itself in lower \fr. As $P$ increases in \TheSystem, we also have to skip more frames to keep the latency in control.

We also studied the effect of other parameter choices of the system in Appendix~\ref{sec:eval_parameter}. Specifically, we demonstrate that as $\lambda$ (the percentile we want to match in \S\ref{sec:design tradeoff}) increases, the video bitrate and the latency decrease. We also show that increasing the pacer pause threshold ($\tau$) does not affect the video bitrate, but the \fr and P95 latency increase.

\section{Related Work}
\label{sec:related}
\NewPara{Congestion Control}.
End-to-end congestion control approaches can be broadly categorized into delay-based~\cite{vegas,arun2018copa,dx,timely,fasttcp,verus,gcc,cardwell2016bbr} or buffer-filling schemes~\cite{pcc,tcp}. Delay-based protocols aim to minimize queuing by adjusting their sending rate based on queuing delay~\cite{shalunov2012low,fasttcp,vegas}, or delay-gradients~\cite{gcc,verus,timely}. 
Buffer-filling algorithms~\cite{pcc,cubic,newreno} send as much traffic as possible until loss or congestion is detected. 
Some approaches like Nimbus~\cite{nimbus} switch between delay-based and buffer-filling modes to improve fairness against competing traffic while maintaining high utilization. However, limited attention has been paid to congestion control for application-limited flows~\cite{rfc7661,jana_cubic_bug} like video traffic that is generated at fixed intervals determined by the \fr.

\NewPara{WebRTC Systems.}
Many video applications 
use Web Real-time Communication (WebRTC)~\cite{webrtc} to deliver real-time video. 
GCC~\cite{carlucci2016analysis}, WebRTC's rate control, uses delay gradients to adjust the sending rate.
However, GCC's conservative behavior coupled with the variance in encoder output results in either under-utilization or latency spikes. 
Salsify~\cite{salsify} previously observed 
 a mismatch between video encoder output and available capacity, and rectified it by encoding multiple versions of the same frame and picking the better match. This requires changing the video codec at the sender and the receiver, making it 
 hard to deploy.
\TheSystem instead matches encoder output to network capacity without changes to the encoder. 
Adaptive bitrate algorithms~\cite{pensieve,buffer-based,mpc,bola,puffer,onrl, zhou2021deadline} solve a similar problem for on-demand video using information about available bandwidth, buffer size, and current bitrate to determine the encoder's target bitrate. 
A recent proposal called SQP~\cite{ray2022sqp}~\footnote{Source code was not available to compare against. \label{no_code}} achieves low end-to-end frame delay for interactive
video streaming applications but operates in much higher bitrates than \TheSystem is designed for. 
\section{Conclusion}
\label{sec:conclusion}

This paper proposes \TheSystem, a new rate control mechanism for low-latency video applications that adapts rapidly to changing network conditions without modifications to the video encoder and the user's hardware.
\TheSystem injects ``dummy'' traffic to make video traffic appear like a backlogged flow running a delay-based congestion controller.
\TheSystem also continuously monitors the frame latencies and adapts the encoder's target bitrate to reduce discrepancies between the encoder output bitrate and link rate. Our evaluation demonstrates its performance gains in enhancing video bitrate and quality while adeptly managing latency.

\label{beforerefs} 


\label{beforerefs}


\newpage
\appendix
\noindent

\section{Probing Mechanisms}
\label{sec:probing_appendix}
\begin{figure}
\centering
    \includegraphics[width=0.75\linewidth]{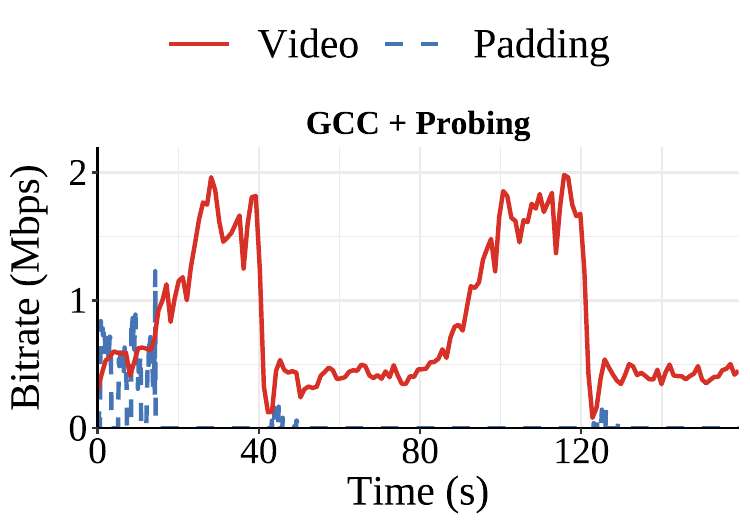}
    \caption{\small Ad-hoc Probing Behavior in WebRTC on a periodic link that alternates between \SI{2}{Mbps} and \SI{500}{Kbps} every 40s. Padding traffic is used when GCC's estimate severely drops (t=40s) but not used when needed during the capacity increase (t=80s).  
    }
\vspace{-17pt}
\label{fig:probing}
\end{figure}

\Fig{probing} displays the performance of probing mechanism within GCC~\cite{webrtc-probing}. This probing controller typically sends probes (padding) in three scenarios: 1) When initiating a connection or after a significant idle period. 2) Upon detecting a potential increase in available bandwidth. 3) Periodically, as part of maintaining an up-to-date estimate of the network conditions, especially when dynamic changes in the network are suspected. We tested this probing mechanism within Google's WebRTC on a link alternating between \SI{2}{Mbps} and \SI{500}{Kbps} every \SI{40}{seconds}. The padding traffic is only sent when GCC significantly reduces the video bitrate (t=40s), but it does not help with bandwidth discovery (t=80s) when the link rate increases. This approach to sending the padding traffic has essentially not provided much benefit for GCC.

\section{Effect of Other Parameter Choices}
\label{sec:eval_parameter}

\begin{figure}[t]
 \centering
 \begin{subfigure}[b]{\linewidth}
     \centering
     \includegraphics[width=\linewidth]{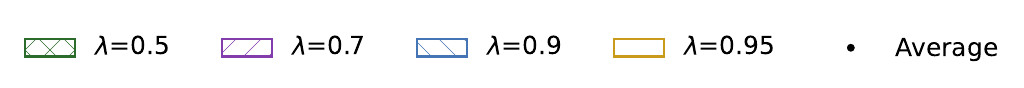}
     \label{fig:lambda_legend}
 \vspace{-18pt}
 \end{subfigure}
 \begin{subfigure}[b]{0.47\linewidth}
     \centering
     \includegraphics[width=\linewidth]{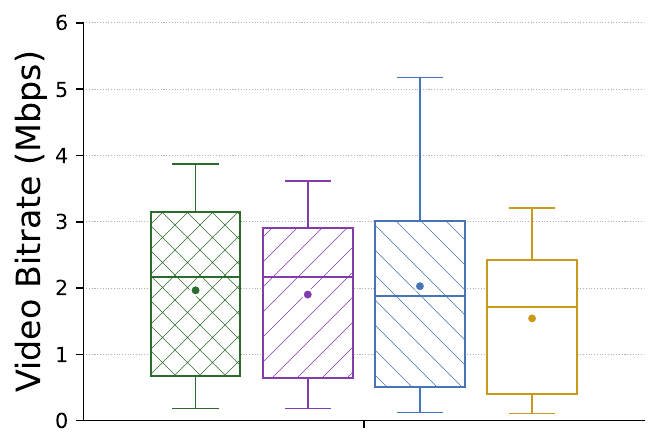}
     \caption{\small Video Bitrate}
     \label{fig:lambda_summary_video}
 \end{subfigure}
 \quad
 \begin{subfigure}[b]{0.47\linewidth}
     \centering
     \includegraphics[width=\linewidth]{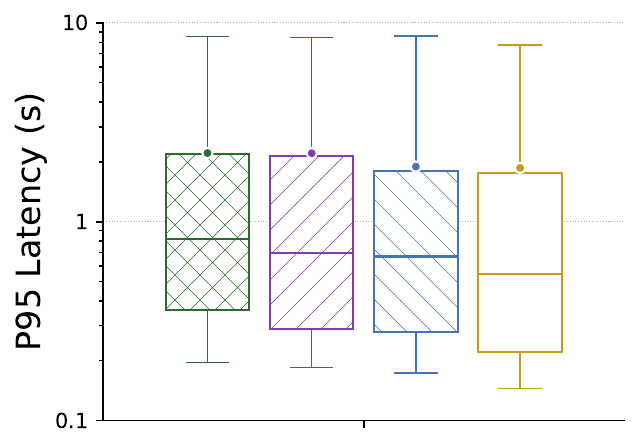}
     \caption{\small Log P95 Latency}
     \label{fig:lambda_summary_latency}
 \end{subfigure}
 \vspace{10pt}
 \begin{subfigure}[b]{0.47\linewidth}
     \centering
     \includegraphics[width=\linewidth]{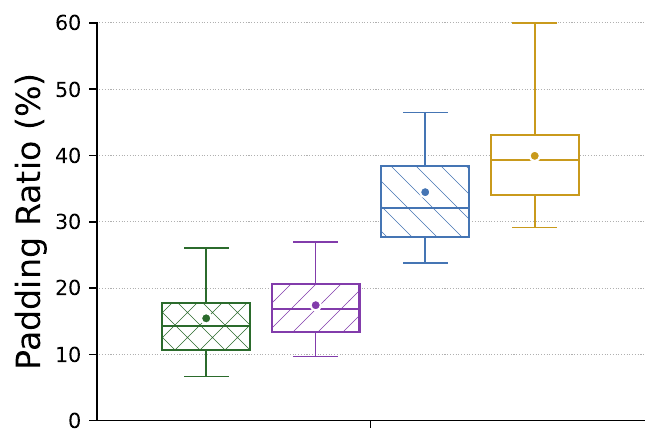}
     \caption{\small Padding Ratio}
     \label{fig:lambda_summary_padding}
 \end{subfigure}
 \quad
 \begin{subfigure}[b]{0.47\linewidth}
     \centering
     \includegraphics[width=\linewidth]{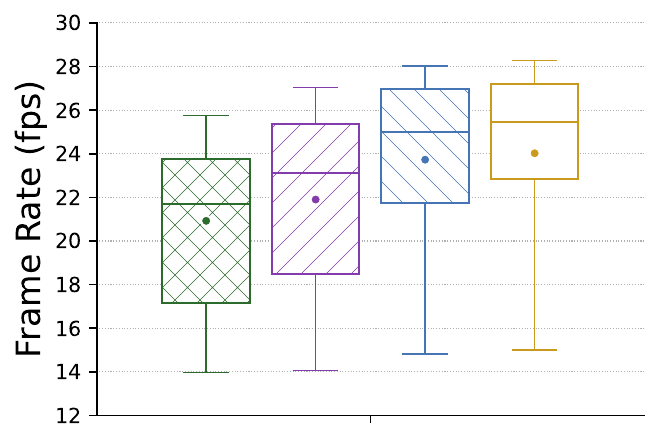}
     \caption{\small Frame Rate}
     \label{fig:lambda_summary_frame_rate}
 \end{subfigure}
 \vspace{-10 pt}
 \caption{\small Effect of $\lambda$ on \TheSystem's performance. Increasing the value of $\lambda$ increases the \fr and padding ratio and decreases the video bitrate and latency on fluctuating links. The whiskers are P5 and P95, the interquartile range shows P25, P50, and P75.}
 \label{fig:lambda_overview}
 \vspace{-20 pt}
\end{figure}

\NewPara{Effect of $\lambda$.}
We evaluate the impact of the parameter $\lambda$, which dictates what fraction of the \fsts we want to tightly bound on all traces and 5 videos. \Fig{lambda_overview} shows the distribution of the metrics across all the videos and traces with $\lambda=0.5, 0.7, 0.9, 0.95$. When $\lambda$ increases, the bitrate selection in \Sec{sec:design tradeoff} is stricter on the \fst values and picks more conservative ones. Hence, the video bitrate decreases (\Fig{lambda_summary_video}), and the latency decreases as well (\Fig{lambda_summary_latency}). Since \TheSystem uses dummy traffic, changes in the video bitrate do not affect \cca estimations and consequently do not change the overall link utilization. Therefore, increasing $\lambda$ increases the padding ratio. The average \fr increases (\Fig{lambda_summary_frame_rate}) as more frames are permitted since their latency is below the threshold.

\begin{figure}[t]
 \centering
 \begin{subfigure}[b]{\linewidth}
     \centering
     \includegraphics[width=\linewidth]{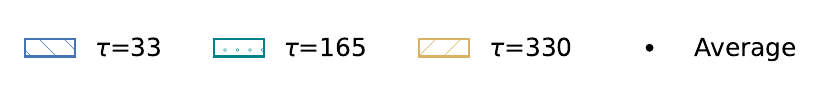}
     \label{fig:kMaxExpectedQueueLength_legend}
 \vspace{-18pt}
  \end{subfigure}
 \begin{subfigure}[b]{0.47\linewidth}
     \centering
     \includegraphics[width=\linewidth]{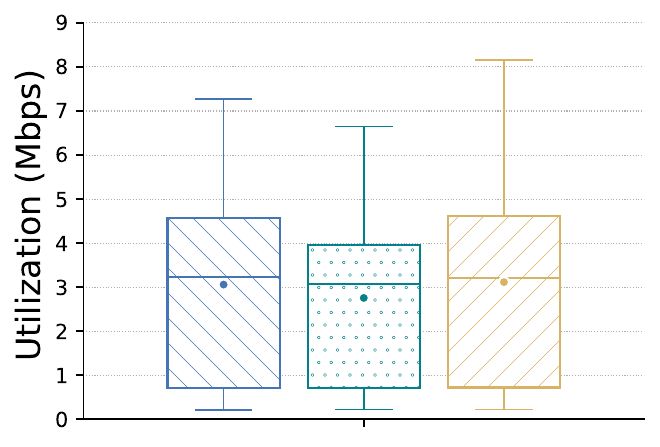}
     \caption{\small Link Utilization}
     \label{fig:kMaxExpectedQueueLength_summary_util}
 \end{subfigure}
  \quad
 \begin{subfigure}[b]{0.47\linewidth}
     \centering
     \includegraphics[width=\linewidth]{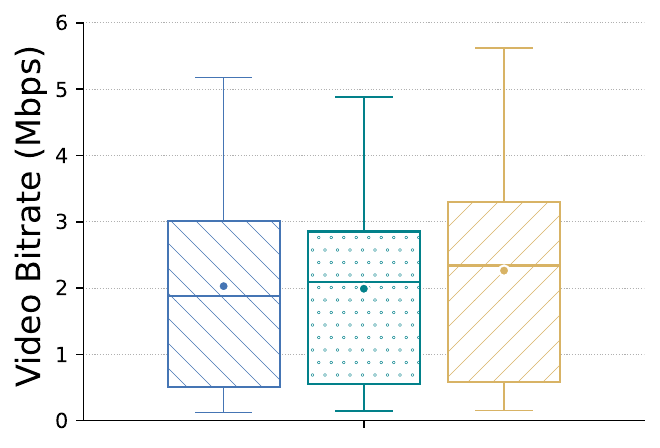}
     \caption{\small  Video Bitrate}
     \label{fig:kMaxExpectedQueueLength_summary_video}
 \end{subfigure}

 \begin{subfigure}[b]{0.47\linewidth}
     \centering
     \includegraphics[width=\linewidth]{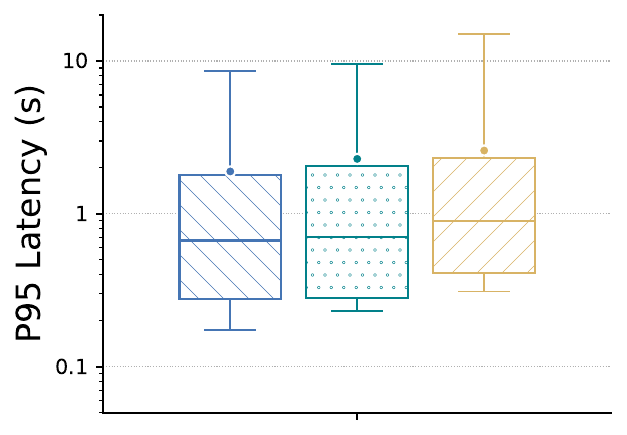}
     \caption{\small Log P95 Latency}
     \label{fig:kMaxExpectedQueueLength_summary_latency}
 \end{subfigure}
 \quad
 \begin{subfigure}[b]{0.47\linewidth}
     \centering
     \includegraphics[width=\linewidth]{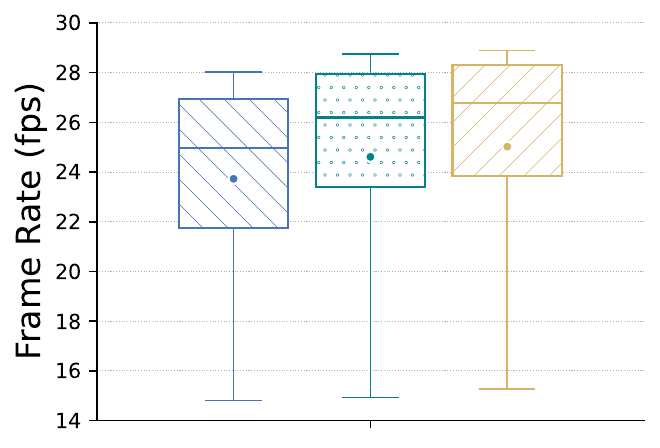}
     \caption{\small Frame Rate}
     \label{fig:kMaxExpectedQueueLength_summary_frame_rate}
 \end{subfigure}
 \caption{\small Effect of pacer queue pause threshold ($\tau$) on \TheSystem. As $\tau$ increases, the \fr increases because fewer frames are skipped, consequently, P95 latency increases. The video bitrate does not change much since the encoder bitrate selector keeps the \fst under control.
 The whiskers are P5 and P95, the interquartile range shows P25, P50, and P75.}
 \label{fig:kMaxExpectedQueueLength_overview}
 \vspace{-15 pt}
\end{figure}

\NewPara{Pacer Queue Pause Threshold ($\tau$).} \Fig{kMaxExpectedQueueLength_overview} shows how the pacer queue pause threshold, $\tau$ (\Sec{sec:design latency}), affects the performance of \TheSystem. \TheSystem skips the current frame if the previous frame has been more than $\tau$ in the pacer queue. We tested \TheSystem with $\tau =$ 33, 165, 330 ms. Changing $\tau$ does not change the network utilization (\Fig{kMaxExpectedQueueLength_summary_util}) because dummy traffic decouples congestion control from the encoder, padding any encoder output to match the link rate. 
The encoder bitrate selection logic computes the target bitrate based on the actual \fst values, so changing $\tau$ does not affect the video bitrate much (\Fig{kMaxExpectedQueueLength_summary_video}) as \fst does not count for the queueing delay frames experience before they're sent.
As $\tau$ increases, the \fr increases (\Fig{kMaxExpectedQueueLength_summary_frame_rate}) because fewer frames are skipped, causing an increase in the P95 latency (\Fig{kMaxExpectedQueueLength_summary_latency}).

\noindent \TheSystem selects $\tau=33 ms$ as it has low latency and relatively high \fr and when compared to GCC.

\section{Report of more Metrics}
\label{sec:more_metrics}
\begin{figure}
    \centering
    \includegraphics[width=0.9\linewidth]{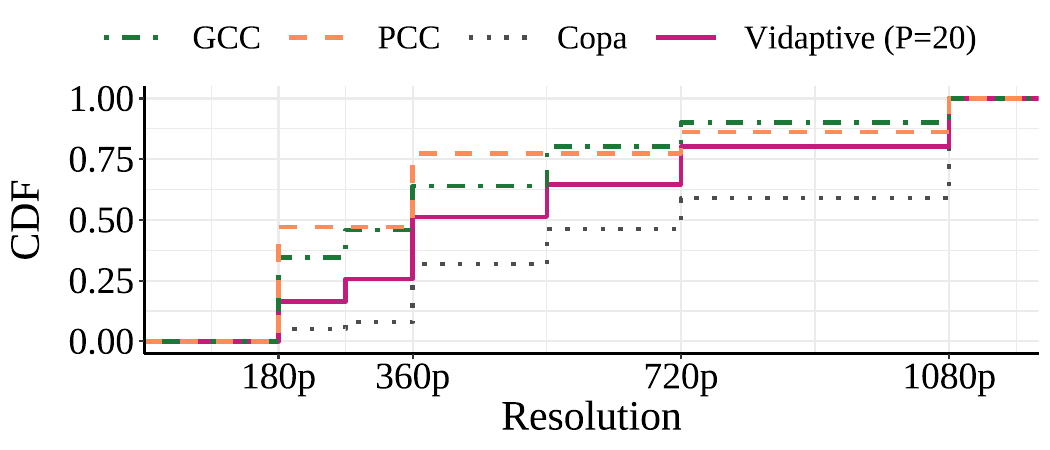}
    \caption{\small CDF of frame resolutions across all the videos and cellular traces. \TheSystem selects higher resolutions in all percentiles.}
    \label{fig:resolution_cdf}
\end{figure}

\NewPara{Cellular Traces.} \Fig{resolution_cdf} shows the CDF of all the selected resolutions during the experiment across all the traces and videos. \TheSystem chooses a higher resolution than GCC and PCC in all the percentiles, which translates to higher video quality. Based on \ref{sec:eval_overall}.

\begin{figure}[t]
 \centering
 \begin{subfigure}[b]{\linewidth}
   \vspace{-5 pt}
     \centering
     \includegraphics[width=\linewidth]{figures/results/eval_overall/overall_legend.pdf}
      \vspace{-10pt}
     \label{fig:cdf_legend}
 \end{subfigure}

 \begin{subfigure}[b]{\linewidth}
     \centering
     \includegraphics[width=\linewidth]{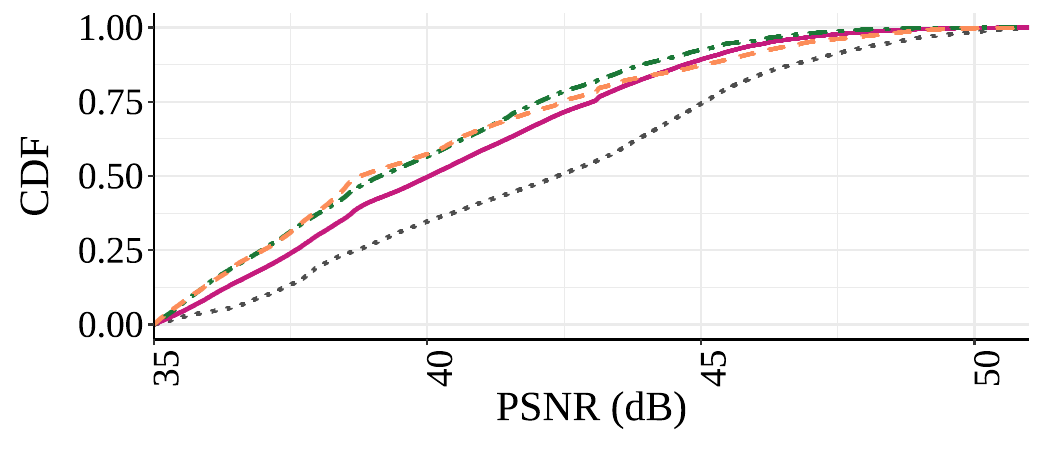}
     \caption{\small PSNR CDF }
     \label{fig:overall_cdf_psnr}
 \end{subfigure}
 \caption{\small CDF of frame PSNR across all the 1M 
 frames.}
 \label{fig:overall_cdfs_appendix}
 \vspace{-10 pt}
\end{figure}

\Fig{overall_cdf_psnr} shows the CDF PSNR for all of the frames across all the traces and all the videos. PSNR follows the same trends as the other visual metrics (SSIM and VMAF). \TheSystem achieves a median PSNR of 38.8 dB compared to GCC's 37.5 dB and PCC's 37.4 dB, and an average of 39.5 dB compared to GCC's 38.2 dB and PCC's 38.1 bB. 

Copa's video bitrate, resolution, and quality are much higher at the cost of significantly longer latencies, such that it practically makes it unsuitable for real-time applications~\cite{itu}. 

\begin{figure}
\centering
    \vspace{-5pt}
    \includegraphics[width=0.9\linewidth]{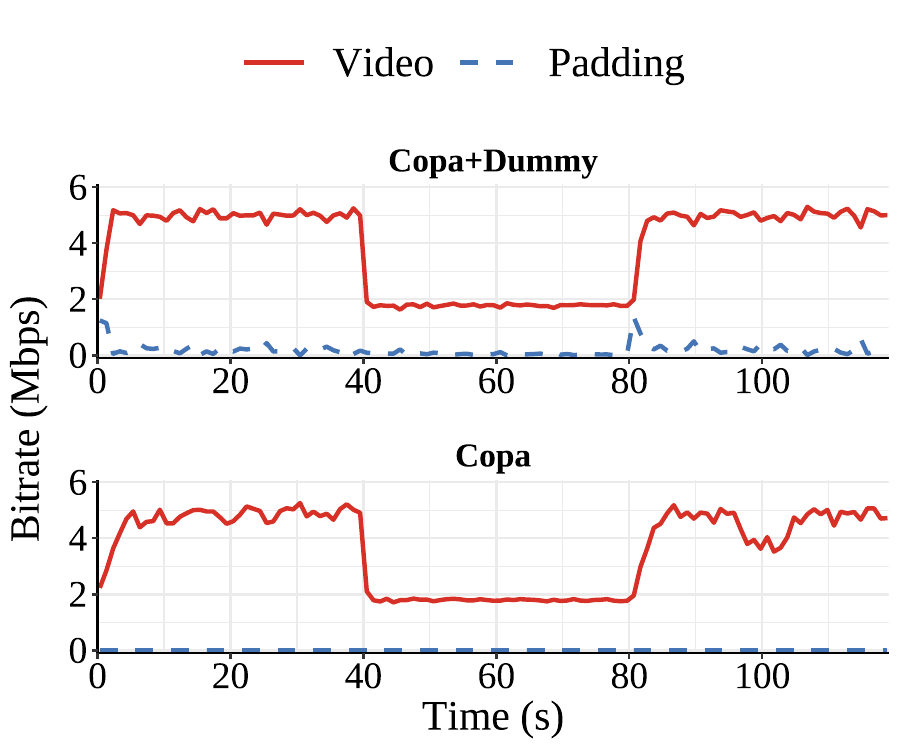}
    \vspace{-5pt}
    \caption{\small Copa with dummy traffic exhibits faster convergence of the video bitrate to the available network capacity
    and maintains a smoother steady-state video bitrate.}
    \vspace{-15pt}
\label{fig:convergence}
\end{figure}

\begin{figure}[t]
 \centering
 \begin{subfigure}[b]{\linewidth}
     \centering
     \includegraphics[width=\linewidth]{figures/results/eval_pantheon/pantheon_legend.pdf}
      \vspace{-10pt}
     \label{fig:cdf_legend}
 \end{subfigure}

 \begin{subfigure}[b]{0.47\linewidth}
     \centering
     \includegraphics[width=\linewidth]{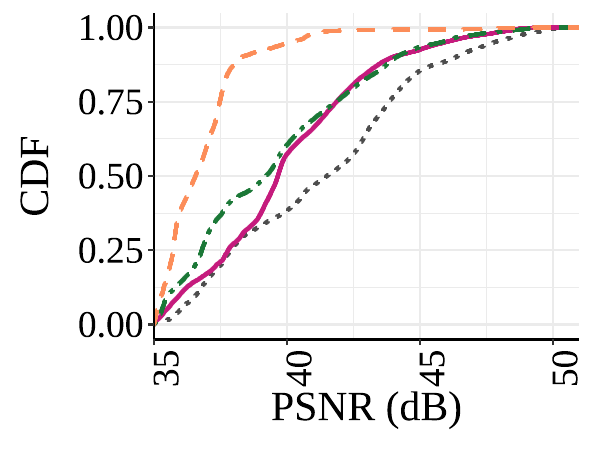}
     \caption{\small PSNR CDF }
     \label{fig:pantheon_cdf_psnr}
 \end{subfigure}
 \hfill
 \begin{subfigure}[b]{0.47\linewidth}
  \vspace{-5pt}
     \centering
     \includegraphics[width=\linewidth]{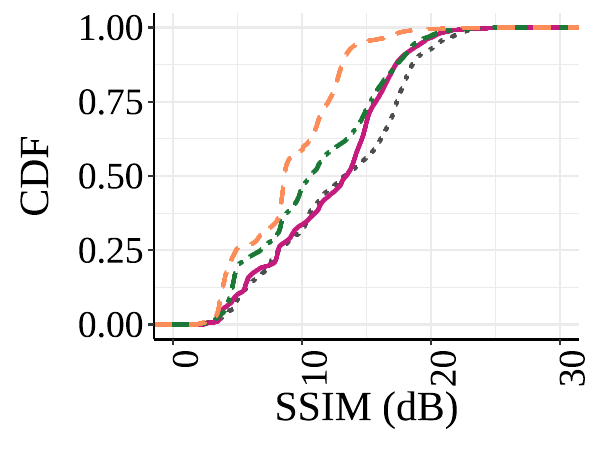}
     \caption{\small SSIM CDF}
     \label{fig:cdf_ssim}
 \end{subfigure}

 \caption{\small CDF of frame PSNR across all the frames for calibrated Pantheon traces.}
 \label{fig:pantheon_cdfs_appendix}
 \vspace{-10 pt}
\end{figure}

\NewPara{Pantheon’s Calibrated Emulators.} \Fig{pantheon_cdfs_appendix} shows the CDF of PSNR and SSIM across all the frames. Compared to PCC, \TheSystem acheives better PSNR and SSIM across all the percentiles. Compared to GCC, \TheSystem improves the median SSIM (\SI{18.8}{dB} $\rightarrow$ \SI{19.4}{dB}) and the median PSNR (\SI{44.5}{dB} $\rightarrow$ \SI{45}{dB}) and the average SSIM (\SI{11}{dB} $\rightarrow$ \SI{12.2}{dB}) and the avergage PSNR (\SI{36.9}{dB} $\rightarrow$ \SI{37.7}{dB}). \TheSystem has lower quality than Copa because Copa achieves higher quality metrics by sacrificing the all percentiles of latency.

\section{Example of Effect of Dummy Traffic}
\label{eval_convergence}
To quantify the effect of dummy traffic in more detail, we disable the changes we made to the target bitrate selection logic and focus on transport layer changes (\Sec{sec:design transport}). In \Fig{convergence}, we emulate a link that starts with \SI{5}{Mbps} of bandwidth for \SI{40}{s}, drops to \SI{2}{Mbps} for the next \SI{40}{s} before jumping back to \SI{5}{Mbps}. 
We compare the video and padding bitrate for ``Copa'' to Copa with dummy traffic (``Copa+Dummy''). Copa takes \SI{6}{s} to match the link rate, while Copa+Dummy takes \SI{2}{s}. The dummy traffic is sent only when the video traffic cannot match the link capacity when it suddenly opens up (around 0s and 80s).
Copa does not match capacity as fast because its rate on the wire is determined by the slow-reacting encoder
(\Sec{sec:motivation}). Further, ``Copa+Dummy'' has a more stable steady-state bitrate than ``Copa'' because the dummy traffic decouples the \cca's feedback from the video encoder's variable output, enabling more accurate link capacity estimation.

\section{Videos}

\NewPara{Dataset Information}
\Tab{dataset_info} summarizes the information of all the videos that we used in the experiments. The videos are all collected from YouTube and cover a different range of motions and settings.

\begin{table}[t] 
    \vspace{15pt}
    \centering
    \small
    \resizebox{\linewidth}{!}{
    \begin{tabular}{lcc}
    \toprule
    \textbf{Video URL} & \textbf{Category} & \textbf{Avg. Bitrate (Kbps)}  \\
    \midrule
\href{https://www.youtube.com/watch?v=6bNOnXTe4Ok}{Video 1} & Ice Skating & 4127\\
\href{https://www.youtube.com/watch?v=Zv11L-ZfrSg}{Video 2} & Wild Animals Collection & 4883\\
\href{https://www.youtube.com/watch?v=sLXWHiIbcIw}{Video 3} & High-motion Cooking & 7278\\
\href{https://www.youtube.com/watch?v=8n1OU9p3fhM}{Video 4} & YouTube Vlogger & 3025 \\
\href{https://www.youtube.com/watch?v=vgezcnqfkdM}{Video 5} & Animated Movie & 2352\\
\href{https://www.youtube.com/watch?v=epfpBdNDljk}{Video 6} & Videocall Stock Footage & 1751\\
\href{https://www.youtube.com/watch?v=ULRlh8kAGf8}{Video 7} & Press Conference & 1410\\
\href{https://www.youtube.com/watch?v=jpvzpzBYwEk}{Video 8} & YouTube Vlogger & 1152\\
\href{https://www.youtube.com/watch?v=OPf0YbXqDm0}{Video 9} & Music Video & 4227\\
\href{https://www.youtube.com/watch?v=lp-EO5I60KA}{Video 10} & Music Video & 4144\\
\href{https://www.youtube.com/watch?v=1XfV40xRq2o}{Video 11} & YouTube Vlogger & 2199\\
\href{https://www.youtube.com/watch?v=hxaFQOac6kM}{Video 12} & YouTube Vlogger & 1379\\
\href{https://www.youtube.com/watch?v=wk9L1N9bRRE}{Video 13} & Art & 3636\\
\href{https://www.youtube.com/watch?v=bRX2MqdsrNw}{Video 14} & Low-motion Cooking & 2502\\
\href{https://www.youtube.com/watch?v=vMEGR3sf3qo}{Video 15} & Football & 6256\\
\href{https://www.youtube.com/watch?v=FDQ-sDDqWvk}{Video 16} & Crafts& 2532\\
\href{https://www.youtube.com/watch?v=hkmnhcsvueE}{Video 17} & Press Conference & 1847 \\
\href{https://www.youtube.com/watch?v=KovN7WKI9Y0}{Video 18} & Screen Sharing / Education & 40 \\
\href{https://www.youtube.com/watch?v=pfBkzWNQTpI}{Video 19} &  Live Music & 4184\\
\href{https://www.youtube.com/watch?v=z2vy1is3aY4}{Video 20} & Weather Forecast & 4754\\
    \bottomrule
    \end{tabular}}
    \caption{Details of our dataset. All videos are at 1920$\times$1080 at 30 \fps and are 4 min long.}
    \label{tab:dataset_info}
    \vspace{-10pt}
\end{table}
\end{document}